\pdfoutput=1
\documentclass[a4paper, 11pt, oneside]{article}

\usepackage{epsf,amsmath,amssymb,graphicx,dcolumn}
\usepackage{scalefnt,cite,hyperref,bbold}
\usepackage{cleveref,etoolbox,color,soul}
\usepackage{listings,booktabs}
\usepackage{epstopdf}
\usepackage{amsfonts}

\newcommand{\half}{\tfrac{1}{2}}
\newcommand{\nc}{\newcommand}

\newlength{\absize}

\nc{\beq}{\begin{equation}}  \nc{\eeq}{\end{equation}}
\nc{\bea}{\begin{eqnarray}}  \nc{\eea}{\end{eqnarray}}

\newcounter{notecount}
\begin{document}

\long\def\symbolfootnote[#1]#2{\begingroup%
\def\thefootnote{\fnsymbol{footnote}}\footnote[#1]{#2}\endgroup}

\title{
\Large\bf\boldmath
One-loop pseudoscalar mass in a 2HDM with a $Z_3$ symmetry
\unboldmath
}

\author{
  \addtocounter{footnote}{2}
  P.~M.~Ferreira$^{(1,2)}$\thanks{\tt pmmferreira@fc.ul.pt}
, Tomás F. Pinto$^{2}$\thanks{\tt tomasfpinto@gmail.com}
\\[0.4cm]
$^{(1)}\!$
  \small Instituto~Superior~de~Engenharia~de~Lisboa~---~ISEL,
  1959-007~Lisboa, Portugal
  \\[2mm]
  $^{(2)}\!$
  \small Centro~de~F\'\i sica~Te\'orica~e~Computacional,
  Faculdade~de~Ci\^encias, Universidade~de~Lisboa
  }
\maketitle

\begin{abstract}
\noindent
A two-Higgs doublet model with a discrete $Z_3$ symmetry acquires, in its scalar and gauge sectors,
an accidental continuous $U(1)$ symmetry. One therefore finds, after spontaneous symmetry breaking
of those symmetries, that a massless pseudoscalar arises, as expected by Goldstone's theorem. In the fermion sector,
however, it is possible to obtain Yukawa matrix textures which distinguish between the $Z_3$ and $U(1)$
symmetries, so one expects loop corrections to originate a non-zero pseudoscalar mass for the $Z_3$ case. We
perform an explicit calculation that shows that at one-loop the pseudoscalar remains massless
for all but one tested $Z_3$ Yukawa matrices. The pseudoscalar mass thus found is highly suppressed by
the hierarchy of fermion masses.
\end{abstract}

\setcounter{footnote}{0}

\section{Introduction}

The Higgs boson was discovered by the LHC collaborations in 2012~\cite{ATLAS:2012yve,CMS:2012qbp} and precision measurements of its properties
have shown it behaves very much as one would expect in the context of the Standard Model (SM)\cite{ATLAS:2022vkf,CMS:2022dwd}. This
was the last particle predicted for the SM that remained to be discovered, and that theory is now ``complete" -- except that
the SM leaves many unanswered questions, such as any explanation for the hierarchy of fermion masses; a satisfactory
description of the matter-antimatter asymmetry in the Universe; and a valid candidate for Dark Matter, among other issues.
Beyond the Standard Model (BSM) theories have been proposed since the SM itself was proposed, and extensions of the
scalar sector, wherein more scalars than just the SM Higgs doublet are considered, have long been a popular and fertile
field of study. Considering an extra scalar gauge singlet has been proposed to allow for an explanation of electroweak
baryogenesis~\cite{Barger:2008jx}; adding a $SU(2)$ triplet to the SM~\cite{Mohapatra:1979ia} originates the see-saw
mechanism, which provides a natural explanation for the smallness of the neutrino masses; and one of the most
simple SM extensions is the two Higgs doublet model (2HDM), proposed by T. D. Lee in 1973~\cite{Lee:1973iz} so
that CP violation could be the result of spontaneous symmetry breaking (for a review, see~\cite{Branco:2011iw}). The
model has a rich phenomenology, predicting the existence of 3 neutral spin-0 particles (not just one) and a charged one.
Some versions of the 2HDM also include natural candidates for dark
matter~\cite{Deshpande:1977rw,Barbieri:2006dq,LopezHonorez:2006gr,Cao:2007rm}. However, while the SM scalar potential is characterized by
2 independent parameters, the most general 2HDM has, seemingly, 14 parameters, though the freedom to redefine
both doublets reduces that number to 11~\cite{Davidson:2005cw}. As such, it is quite useful to impose discrete or continuous
global symmetries on the 2HDM, thereby reducing the number of parameters of the model and increasing its predictive power.
Not only that but those symmetries can provide natural solutions for phenomenological problems: for instance, a $Z_2$ symmetry
was proposed by Glashow, Weinberg and Paschos~\cite{Glashow:1976nt,Paschos:1976ay} so that flavour changing neutral currents (FCNCs)
mediated by neutral scalars become forbidden -- and since such interactions are present in the most general 2HDM lagrangian, and
are highly constrained by experimental data, the $Z_2$ symmetry provides a natural explanation for their absence, all the while
reducing the number of free parameters in the scalar potential to 7. Likewise, a continuous $U(1)$ symmetry was proposed by Peccei
and Quinn as a possible explanation for the strong QCD problem~\cite{Peccei:1977hh}, from which a 2HDM scalar potential with only 6
parameters emerges.

All in all, six different global symmetries may be imposed on the $SU(2)_L \times U(1)_Y$ scalar
potential~\cite{Ivanov:2006yq,Ivanov:2007de}. Due to the freedom to redefine doublets by choosing different basis for those fields,
two {\em a priori} different scalar potentials may well correspond to the same symmetry. For instance, the $Z_2$ symmetry, resulting
from the field transformations $\Phi_1\rightarrow \Phi_1$ and $\Phi_2\rightarrow -\Phi_2$, yields a scalar potential with the same
physics than a potential invariant under a permutation symmetry, $\Phi_1\leftrightarrow \Phi_2$, even though both potentials
have very different relations imposed on its parameters. Another issue is that of {\em accidental symmetries}: since we are
considering a renormalizable theory the scalar potential will have at most quartic terms in the fields, different field transformations
yield physically equivalent potentials. For instance, a scalar potential invariant under a $Z_3$ symmetry, where doublets transform
as $\Phi_1\rightarrow \Phi_1$ and $\Phi_2\rightarrow e^{2i \pi/3} \Phi_2$, is physically indistinguishable from one invariant
under a $U(1)$ Peccei-Quinn symmetry, for which  $\Phi_1\rightarrow \Phi_1$ and $\Phi_2\rightarrow e^{i \theta} \Phi_2$, with
arbitrary values for $\theta$. Indeed, while we consider only the scalar and gauge sectors, there is no distinction between
these two possibilities. The same can be said for a 2HDM scalar sector invariant under a $Z_N$ symmetry, with $N\geq 3$,
and as mentioned this is caused by the fact that the scalar potential only has terms up to fourth power in the doublet fields.
However, this picture changes when one considers the Yukawa sector. Though most of the six symmetries of the scalar potential can
be extended to the full lagrangian yielding acceptable fermion phenomenology, there are different possibilities of achieving that.
We can therefore have physically different theories which are identical in the scalar and gauge sectors but differ substantially
in their Yukawa interactions. For instance, it is possible to have a $U(1)$-invariant 2HDM where its Yukawa sector has no
FCNCs, and a $Z_3$ symmetric 2HDM, which shares the same scalar potential as the $U(1)$ theory, but with FCNCs in its scalar-fermion
interactions.

This then raises an interesting question, which we will study in this paper: since the $U(1)$ 2HDM is invariant under a continuous symmetry,
any vacuum which spontaneously breaks that symmetry ({\em i.e.} one for which both doublets acquire a vacuum expectation value (vev))
yields a massless scalar (other than the 3 Goldstone bosons one expects from electroweak symmetry breaking)- Indeed, minimising the
scalar potential and computing the scalar masses one finds that the pseudoscalar of the model, $A$, is massless in this
theory~\footnote{This is usually avoided by adding a soft-breaking quadratic term in the potential. In this work we will only
deal with potentials without such soft breakings. It should be noted, however, that the Peccei-Quinn 2HDM symmetry is anomalous, and a 
non-zero but very small axion mass is expected to appear~\cite{Srednicki:1985xd} from, in principle, 3-loop contributions. For 
an alternate way to generate axion masses in 2HDMs, see~\cite{BhupalDev:2014bir,Darvishi:2019ltl}.}.
This is to be expected from Goldstone's theorem\cite{Goldstone:1961eq,Goldstone:1962es}. But with a $Z_3$ 2HDM we are left with a
lagrangian which is invariant under a {\em discrete} symmetry but whose scalar potential is identical to the Peccei-Quinn one. One then
finds at tree-level that upon spontaneous symmetry breaking the pseudoscalar $A$ is also massless in this case. Is this a possible exception of Goldstone's
theorem, which explicitly applies to models with continuous symmetries? We will show in the present work that that is not the case and
Goldstone's theorem remains valid. Since both models, the $U(1)$ and $Z_3$ invariant 2HDMs, differ only in their Yukawa sectors, we should compute
the contributions from fermions to scalar masses, and hopefully see that those mass corrections to the pseudoscalar mass are non-zero
for the $Z_3$ model, remaining zero for the $U(1)$ case. But fermions have no contribution to scalar masses at tree-level, which leads us to
compute one-loop corrections to the scalar sector.  We will discover that indeed a $Z_3$ symmetry yields a massive pseudoscalar at the one-loop
level, while all $U(1)$-invariant Yukawa sectors still originate a massless $A$. However, we will also show that, out of all possible $Z_3$-invariant
Yukawa matrices (the seemingly complete list of such textures was presented in~\cite{Ferreira:2010ir} ), only {\em one} gives a non-zero mass to $A$.

This paper is organised as follows: in section~\ref{sec:mod} we present the 2HDM and discuss in detail the $U(1)$ and $Z_3$ symmetries imposed
on it. We show that at tree-level both symmetries yield a massless pseudoscalar. We review how these symmetries are extended to the fermion
sector, and how they generate different Yukawa textures. In section~\ref{sec:onel} we present the one-loop potential and discuss its minimisation
and the formalism needed to compute one-loop contributions to the masses of the scalars of the model. We present the contributions to the
pseudoscalar and neutral Goldstone masses from the scalar and gauge sectors in section~\ref{sec:onelsg} and show how both of those
masses remain zero, even at one-loop. In section~\ref{sec:onelf} we then compute the one-loop corrections from fermions,
showing that a specific $Z_3$ Yukawa texture yields a non-zero pseudoscalar mass, all the while preserving the masslessness of the
neutral Goldstone boson. We will also show that the other possible $Z_3$ Yukawa textures still yield a massless pseudsocalar,
as do all $U(1)$-invariant Yukawas. We will discuss these results and present our conclusions in section~\ref{sec:conc}.

\section{The Two Higgs Doublet Model: $U(1)$ and $Z_3$ symmetries}
\label{sec:mod}

The 2HDM extends the scalar sector of the SM by considering two hypercharge $Y = 1$ $SU(2)_L$ doublets.
The most general renormalizable scalar potential that one can then write which is invariant
under the $SU(2)_L\times U(1)_Y$ gauge symmetries is given by, at tree-level,
\bea
V_0 &=& m_{11}^2\Phi_1^\dagger\Phi_1+m_{22}^2\Phi_2^\dagger\Phi_2
-[m_{12}^2\Phi_1^\dagger\Phi_2+{\rm h.c.}]+\half\lambda_1(\Phi_1^\dagger\Phi_1)^2
+\half\lambda_2(\Phi_2^\dagger\Phi_2)^2
+\lambda_3(\Phi_1^\dagger\Phi_1)(\Phi_2^\dagger\Phi_2)\nonumber\\[8pt]
&&\quad
+\lambda_4(\Phi_1^\dagger\Phi_2)(\Phi_2^\dagger\Phi_1)
+\left\{\half\lambda_5(\Phi_1^\dagger\Phi_2)^2
+\big[\lambda_6(\Phi_1^\dagger\Phi_1)
+\lambda_7(\Phi_2^\dagger\Phi_2)\big]
\Phi_1^\dagger\Phi_2+{\rm h.c.}\right\}\,.
 \label{eq:pot}
\eea
Other than the $m^2_{12}$ and $\lambda_{5,6,7}$ coefficients, all parameters are real. Since the
two doublets are not physical fields (those will be the scalar mass eigenstates found after spontaneous symmetry
breaking), any unitary transformation of $\Phi_1$ and $\Phi_2$ which leaves the theory's kinetic terms invariant
leads to a physically equivalent model. These {\em basis transformations}, defined as
\beq
\Phi^\prime_i \,=\, U_{ij} \,\Phi_j\,,
\eeq
for a generic $2\times 2$ unitary matrix $U$, allow one to simplify the model in many circumstances. For instance,
such basis transformations can show that the most general 2HDM scalar potential has only 11 independent
real parameters (and not 14 as a na\"{\i}ve parameter counting of eq.~\eqref{eq:pot} would lead to conclude~\cite{Davidson:2005cw}).

The most general 2HDM, when extended to the fermion sector, would induce tree-level flavour-changing neutral currents (FCNC)
mediated by scalar particles. These interactions may have sizeable contributions to several meson physics observables
(such as $B$ and $K$ meson oscillation parameters, for instance) and as such are constrained by experiments to be quite small.
One way to naturally ensure the absence of these FCNC is to impose a discrete symmetry on the lagrangian --- a simple $Z_2$
symmetry, as proposed by Glashow, Weinberg and Paschos~\cite{Glashow:1976nt,Paschos:1976ay}, wherein invariance of the lagrangian
is required for the
transformation $\Phi_1 \rightarrow \Phi_1$, $\Phi_2 \rightarrow -\Phi_2$, forces a single scalar doublet to couple to fermions
of the same electric charge, thereby eliminating tree-level FCNC in scalar-fermion interactions. The impact of this symmetry on
the scalar potential is to set to zero the parameters $m^2_{12}$, $\lambda_6$ and $\lambda_7$~\footnote{In the basis where the $Z_2$ symmetry
has the form mentioned. In other basis there might appear different relations between parameters but the physical consequences
of said symmetry are the same.}.

Another global symmetry that one can impose on the 2HDM is a continuous $U(1)$ symmetry, as proposed by Peccei and Quinn~\cite{Peccei:1977hh}
to solve the strong QCD problem. To wit, one requires invariance of the lagrangian under the transformations
\beq
\Phi_1 \rightarrow \Phi^\prime_1 \,=\, \Phi_1\;\;\; , \;\;\; \Phi_2 \rightarrow \Phi^\prime_2 \,=\, e^{i\,\theta} \Phi_2\, ,
\label{eq:u1}
\eeq
with $\theta$ an arbitrary real number. This induces another restriction on the 2HDM scalar parameters, rendering $\lambda_5$ equal to zero,
so that the potential becomes
\beq
V_0\,=\, m_{11}^2\Phi_1^\dagger\Phi_1+m_{22}^2\Phi_2^\dagger\Phi_2
+\half\lambda_1(\Phi_1^\dagger\Phi_1)^2
+\half\lambda_2(\Phi_2^\dagger\Phi_2)^2
+\lambda_3(\Phi_1^\dagger\Phi_1)(\Phi_2^\dagger\Phi_2)
+\lambda_4(\Phi_1^\dagger\Phi_2)(\Phi_2^\dagger\Phi_1)\,.
 \label{eq:potU1}
\eeq
Let us now consider electroweak symmetry breaking, assuming both doublets acquire real and neutral vevs~\footnote{It is actually impossible
that charge breaking vevs occur for this model. Likewise, complex vevs which would spontaneously break CP cannot occur in this
model as well~\cite{Ferreira:2010hy}.}, $\langle \Phi_1\rangle = ( 0 \, , \,  v_1/\sqrt{2} )^T$ and
$\langle \Phi_2\rangle = (0 \, , \,  v_2/\sqrt{2} )^T$
such that $v_1^2 + v_2^2 = v^2$, with $v = 246$ GeV for correct electroweak symmetry breaking. This vacuum
clearly breaks the global $U(1)$ symmetry imposed on the model~\footnote{For some regions of parameter space, the model also
allows for
a vacuum in which one of the doublets remains vevless and therefore preserves the $U(1)$ symmetry. That yields a Inert
2HDM~\cite{Deshpande:1977rw,Barbieri:2006dq,LopezHonorez:2006gr,Cao:2007rm} and would have a non-zero pseudoscalar mass already at tree-level.}.
We then parameterize the doublets' real components $\varphi_i$ such that
\beq
\Phi_1=
\frac{1}{\sqrt{2}}\begin{pmatrix}
	\varphi_1+i\varphi_3 \\
	v_1\,+\,\varphi_5+i\varphi_7
\end{pmatrix}, \qquad\quad
\Phi_2=
\frac{1}{\sqrt{2}}\begin{pmatrix}
	\varphi_2+i\varphi_4 \\
	v_2\,+\,\varphi_6+i\varphi_8
\end{pmatrix}\,,
\label{eq:doub}
\eeq
so that the tree-level minimization conditions are found by setting all $\varphi_i$ to zero and computing the derivatives of the
potential in order of the vevs,
\bea
\frac{\partial V_0}{\partial v_1} = 0 & \Longleftrightarrow & \left(m^2_{11} \,+\, \frac{1}{2} \lambda_1 v_1^2 + \frac{1}{2} \lambda_{34}  v_2^2 \right) v_1 = 0
\nonumber \\
\frac{\partial V_0}{\partial v_2} = 0 & \Longleftrightarrow & \left(m^2_{22} \,+\, \frac{1}{2} \lambda_2 v_2^2 + \frac{1}{2} \lambda_{34} v_1^2 \right) v_2 = 0\,,
\label{eq:min0}
\eea
where for convenience we defined $\lambda_{34} =\lambda_3 + \lambda_4$.
The tree-level scalar squared mass matrix is then given by
\beq
\left[M^2_S \right]_{ij}\,=\,\left.\frac{\partial^2 V_0}{\partial \varphi_i \partial \varphi_j}\right|_{min}\,
\eeq
where by ``$min$" we indicate that these second derivatives are computed with all $\varphi_i = 0$. With the vevs considered and our ordering
of the real components $\varphi_i$, the $8\times 8$ mass matrix above breaks into four $2\times 2$ blocks. Two of those blocks (corresponding to
$\{\varphi_1, \varphi_2\}$ and $\{\varphi_3, \varphi_4\}$) are identical and give the charged mass matrix,
\beq
\left[M^2_{H^\pm} \right]\,=\,\begin{pmatrix}
m^2_{11} \,+\, \half \lambda_1 v_1^2 + \frac{1}{2} \lambda_3 v_2^2 & \frac{1}{2} \lambda_4 v_1 v_2 \\[8pt]
\frac{1}{2} \lambda_4 v_1 v_2 & m^2_{22} \,+\, \frac{1}{2} \lambda_2 v_2^2 + \frac{1}{2} \lambda_3 v_1^2
\label{eq:charg}
\end{pmatrix}\,.
\eeq
If evaluated at the tree-level solution of the minimization equations~\eqref{eq:min0}, this matrix has a zero eigenvalue, corresponding
to the mass of the charged Goldstone $G^\pm$, and a non-zero one giving the tree-level charged scalar mass. The block corresponding
to fields $\{\varphi_5, \varphi_6\}$ is the CP-even mass matrix,
\beq
\left[M^2_{H} \right]\,=\,\begin{pmatrix}
m^2_{11} \,+\,\tfrac{3}{2} \lambda_1 v_1^2 + \frac{1}{2} \lambda_{34} v_2^2 &  \lambda_{34} v_1 v_2 \\[8pt]
\lambda_{34} v_1 v_2 & m^2_{22} \,+\, \frac{3}{2} \lambda_2 v_2^2 + \frac{1}{2} \lambda_{34} v_1^2
\label{eq:cpeven}
\end{pmatrix}\,.
\eeq
When the eigenvalues of this matrix are evaluated with vevs satisfying the conditions of eqs.~\eqref{eq:min0} we find two non-zero
values, the squared masses of the lighter CP-even scalar, $h$, and the heavier one, $H$.

Finally, the block corresponding to the component fields $\{\varphi_7, \varphi_8\}$, which are the imaginary parts of the
neutral components of both doublets, will give us the masses of the pseudoscalar $A$ and the neutral Goldstone boson $G^0$. This block
is found to be diagonal, and the tree-level pseudoscalar mass matrix is then given by
\beq
\left[M^2_{A} \right]\,=\,\begin{pmatrix}
m^2_{11} \,+\,\tfrac{1}{2} \lambda_1 v_1^2 + \frac{1}{2} \lambda_{34} v_2^2 &  0 \\[8pt]
0 & m^2_{22} \,+\, \frac{1}{2} \lambda_2 v_2^2 + \frac{1}{2} \lambda_{34} v_1^2
\label{eq:pseud}
\end{pmatrix}\,.
\eeq
At the tree-level minimum which satisfies eqs.~\eqref{eq:min0}, we find that all four entries of this matrix vanish -- which means that,
in accordance with Goldstone's theorem, the spontaneous breaking of the continuous Peccei-Quinn $U(1)$ symmetry produces a massless pseudoscalar,
along with the massless neutral Goldstone boson one would expect since electroweak symmetry breaking is also occurring.
As for the gauge boson masses, they arise as usual from the kinetic terms in the lagrangian,
\beq
{\cal L}_K\,=\, (D_\mu \Phi_1)^\dagger  (D^\mu \Phi_1)\,+\,(D_\mu \Phi_2)^\dagger  (D^\mu \Phi_2)
\label{eq:kin}
\eeq
where the covariant derivatives are written as
\beq
D_\mu \,=\, \partial_\mu\,-\, \frac{i}{2}\,g^\prime\, B_\mu\,-\,\frac{i}{2}\, g\, \sigma^a\, W^a_\mu\,,
\eeq
with $a = \{1,2,3\}$ and $\sigma^a$ are the Pauli matrices. The usual 2HDM gauge boson mass matrices
are then found, and the gauge boson masses given by
\beq
m^2_W\,=\,\frac{g^2}{4}\,(v_1^2 + v_2^2)\;\;\; , \;\;\;
m^2_Z\,=\,\frac{g^2 + {g^\prime}^2}{4}\,(v_1^2 + v_2^2)\,.
\label{eq:gauge}
\eeq

At this stage, let us consider the 2HDM with a different symmetry -- a $Z_3$ symmetry, whereas instead of a generic
angle $\theta$ as was considered in the $U(1)$ symmetry in transformation of eq.~\eqref{eq:u1}, we restrict ourselves to
a single complex phase of $2\pi/3$ and require invariance of the lagrangian under
\beq
\Phi_1 \rightarrow \Phi^\prime_1 \,=\, \Phi_1\;\;\; , \;\;\; \Phi_2 \rightarrow \Phi^\prime_2 \,=\, e^{2 i \pi/3} \Phi_2\, .
\label{eq:z3}
\eeq
Now, it is easy to verify that requiring invariance under $Z_3$ or $U(1)$ yields exactly the same scalar potential, eq.~\eqref{eq:potU1}.
This is because all terms of the form $\Phi_1^\dagger \Phi_2$ (or their hermitian conjugates) transform non-trivially under both
symmetries~\footnote{Albeit with different complex phases, $e^{i\theta}$ for $U(1)$ and $e^{2i\pi/3}$ for $Z_3$.} and as such all coefficients
which multiply such a term, with the exception of $\lambda_4$, are set to zero by both symmetries. As a consequence, imposing the discrete $Z_3$ symmetry
on the 2HDM scalar potential results in an {\em accidental} continuous $U(1)$ symmetry, and as such it is not surprising that spontaneous
symmetry breaking yields a massless pseudoscalar in both cases. The gauge sector also does not distinguish between both symmetries so
a scalar + gauge 2HDM with a $Z_3$ symmetry is indeed, by accident, a Peccei-Quinn $U(1)$ 2HDM. Indeed, the same occurs for any other $Z_N$ symmetry,
with $N\geq 3$.

However, this picture changes when we introduce fermions in the theory. We will only look at the quark sector, but leptons
would be handled in a similar manner. The interactions between quarks and scalars are given by the Yukawa lagrangian,
written as
\beq
-{\cal L_Y}\,=\, \bar{Q}_L\Gamma_1 \Phi_1 n_R\,+\,\bar{Q}_L\Gamma_2 \Phi_2 n_R\,+\,\bar{Q}_L\Delta_1 \tilde{\Phi}_1 p_R\,+\,\bar{Q}_L\Delta_2 \tilde{\Phi}_2 p_R
+{\rm h.c.}\,,
\label{eq:yuk}
\eeq
where the $\Gamma_i$, $\Delta_i$ are $3\times 3$ complex matrices of Yukawa couplings, and $n_R$ ($p_R$) are 3-vectors in flavour space
containing the 3 righthanded negative (positive) quark
fields who, upon rotation to the basis of quark mass eigenstates, will describe the down (up) quarks. Finally, $Q_L$ stands for a 3-vector in flavour space
containing the 3 left quark doublets, {\em i.e.} $(Q_L)_i = (p_i \,,\, n_i)_L^T$. The quark mass matrices are then given by
\beq
M_n\,=\,\frac{1}{\sqrt{2}}\,(\Gamma_1 v_1 + \Gamma_2 v_2)\;\;\; ,\;\;\; M_p\,=\,\frac{1}{\sqrt{2}}\,(\Delta_1 v_1 + \Delta_2 v_2)\,,
\label{eq:Mnp}
\eeq
which are diagonalised by $3\times 3$ unitary matrices $U_{qL}$, $U_{qR}$ such that the physical down and up masses appear as
\bea
M_d\,=\,{\rm diag} (m_d\,,\,m_s\,,\,m_b) & = & U_{dL}^\dagger\,M_n\, U_{dR}\,, \nonumber \\
M_u\,=\,{\rm diag} (m_u\,,\,m_c\,,\,m_t) & = & U_{uL}^\dagger\,M_p\, U_{uR}\,.
\label{eq:Mdu}
\eea
Extending a $U(1)$ or $Z_3$ symmetry to the Yukawa sector is achieved by requiring invariance of the lagrangian~\eqref{eq:yuk} under simultaneous
transformations of the scalars under eqs.~\eqref{eq:u1} or \eqref{eq:z3}, respectively, and of the fermion fields under
\bea
Q_L &\rightarrow & Q^\prime_L = {\rm diag} (e^{i\alpha_1}\,,\,e^{i\alpha_2}\,,\,e^{i\alpha_3})\,Q_L \,, \nonumber \\
n_R &\rightarrow & n^\prime_R =  {\rm diag} (e^{i\beta_1}\,,\,e^{i\beta_2}\,,\,e^{i\beta_3})\,n_R \,, \nonumber \\
p_R &\rightarrow & p^\prime_R =  {\rm diag} (e^{i\gamma_1}\,,\,e^{i\gamma_2}\,,\,e^{i\gamma_3})\,p_R \,,
\label{eq:phq}
\eea
with generic real phases $\alpha_i$, $\beta_i$ and $\gamma_i$.
Under these transformations the Yukawa matrices transform as
\begin{eqnarray}
\left( \Gamma_a \right)_{ij}
&=&
e^{i (\alpha_i - \beta_j - \theta_a)} \left( \Gamma_a \right)_{ij},
\label{eq:Gammatr}
\\
\left( \Delta_a \right)_{ij}
&=&
e^{i (\alpha_i - \gamma_j + \theta_a)} \left( \Delta_a \right)_{ij},
\label{eq:Deltatr}
\end{eqnarray}
where $\theta_1 = 0$ and $\theta_2 = \theta$. A given entry of a $\Gamma$ or $\Delta$ matrix will be non-zero if and only if
the combinations of phases in the exponentials above are equal to 0 (mod($2\pi$)).
The complete~\footnote{It should be emphasised that in ref.~\cite{Ferreira:2010ir} only textures resultant from abelian symmetries were
obtained. Other textures are therefore possible. It also does not seem that the $Z_3$ textures considered in~\cite{Ferreira:2011xc}
are included in the list provided in~\cite{Ferreira:2010ir}.} list of possible Yukawa matrices that are (a) invariant under
these transformations, (b) generate 3 massive up and down quarks and (c) can produce a realistic Cabbibo-Kobayashi-Maskawa (CKM)
quark mixing matrix were obtained in ref.~\cite{Ferreira:2010ir}, where it was shown that there are many different, and physically
non-equivalent, ways of extending symmetries of the scalar sector to the Yukawa one. For instance, there are many Yukawa matrix textures
which are themselves $U(1)$ invariant. One example out of many is to choose $\alpha_i = 0$ and $\beta_i = -\gamma_i = -\theta$ for all $i = 1,2,3$,
which forces all fermions to couple only to $\Phi_2$, the Yukawa matrices resulting thereof given by
\bea
\Gamma_1 \, = \, \begin{pmatrix} 0 & 0 & 0 \\ 0 & 0 & 0 \\ 0 & 0 & 0 \end{pmatrix}  & , &
\Gamma_2 \, = \, \begin{pmatrix} \times & \times & \times \\ \times & \times & \times \\ \times & \times & \times \end{pmatrix}\,, \nonumber \\
\Delta_1 \, = \, \begin{pmatrix} 0 & 0 & 0 \\ 0 & 0 & 0 \\ 0 & 0 & 0 \end{pmatrix}  & , &
\Delta_2 \, = \, \begin{pmatrix} \times & \times & \times \\ \times & \times & \times \\ \times & \times & \times \end{pmatrix}\,,
\label{eq:tI}
\eea
where ``$\times$" denotes a generic complex number. This is a well-known realisation of a ``Type I" 2HDM, and the Yukawa matrices
above have no scalar-mediated FCNCs. Like many other $U(1)$-invariant textures, they are obtained for completely generic values of the
phase $\theta$.

But there are Yukawa textures which can only be obtained if $\theta = 2\pi/3$ and are thus specifically $Z_3$-symmetric~\footnote{There are
also Yukawa matrix textures which can only be obtained if $\theta = \pi$ and are therefore $Z_2$-symmetric. They are listed in section III.C.4
of ref.~\cite{Ferreira:2010ir} but we will not consider them in this paper.}. All such cases are listed in section III.C.5
of ref.~\cite{Ferreira:2010ir}~\footnote{We detected a typo in eq. (89) of ref.~\cite{Ferreira:2010ir}: the (1,3) entry of matrix $\Gamma_1$
should be zero.}. One such example are the textures
\bea
\Gamma_1 \, = \, \begin{pmatrix} \times & 0 & 0 \\ 0 & \times & 0 \\ 0 & 0 & \times \end{pmatrix}  & , &
\Gamma_2 \, = \, \begin{pmatrix} 0 & \times & 0 \\ 0 & 0 & \times \\ \times & 0 & 0 \end{pmatrix}\,, \nonumber \\
\Delta_1 \, = \, \begin{pmatrix} 0 & \times & 0 \\ 0 & 0 & \times \\ \times & 0 & 0 \end{pmatrix}  & , &
\Delta_2 \, = \, \begin{pmatrix} \times & 0 & 0 \\ 0 & \times & 0 \\ 0 & 0 & \times \end{pmatrix}  \,.
\label{eq:Yukz3}
\eea
Looking at the entries of the $\Gamma$ matrices we obtain the following conditions on the $\alpha$ and $\beta$
phases (we already chose $\alpha_1 = 0$,  which can always be made without loss of
generality)~\footnote{In these equations obviously all terms equal to ``0" should be interpreted as ``0 (mod($2\pi$))", but for
our purposes it is enough to only consider a factor of ``$2n\pi$" in the last equation.},
\bea
\Gamma_1: & & -\beta_1 = 0 \quad ; \quad \alpha_2 - \beta_2 = 0 \quad ; \quad \alpha_3 - \beta_3 = 0 \nonumber \\
\Gamma_2: & & -\beta_2 - \theta = 0 \quad ; \quad \alpha_2 - \beta_3  - \theta = 0 \quad ; \quad \alpha_3 - \beta_1  - \theta = 2n\pi\,,
\label{eq:fases}
\eea
with $n$ a generic integer number.
It is then simple to verify that for these equations to have a solution one must have $\alpha_2 = \beta_2 = -\theta$ and
$\alpha_3 = \beta_3 = -2\theta$. The final equation then implies $-3\theta = 2n\pi$, for which $\theta = 2\pi/3$ is a solution
(as is $\theta = -2\pi/3$, but both solutions correspond to exactly the same physics).

The models corresponding to the $U(1)$ textures of~\eqref{eq:tI} and the $Z_3$ ones of~\eqref{eq:Yukz3} are not the same: the former has
neutral scalars which preserve flavour in their interactions with quarks, whereas the latter has FCNCs in those same interactions.
There is therefore a physical distinction between the $U(1)$ and $Z_3$ lagrangians, even though their scalar sectors are seemingly identical.
But then we are left with a question: if the $Z_3$ lagrangian has a discrete symmetry, then Goldstone's theorem does not apply, and
there should be a massive pseudoscalar after spontaneous symmetry breaking. That, however, is not what we have already seen in
eq.~\eqref{eq:pseud}, where the tree-level minimisation conditions yield a massless pseudoscalar. Before considering that Godlstone's
theorem might be incorrect, however, we should consider that both lagrangians, the $U(1)$ and $Z_3$-symmetric ones, differ only in their
Yukawa sector, and as such we may anticipate that contributions to the pseudoscalar mass from the fermionic sector will distinguish
between both models, and give rise to a non-zero mass for the $Z_3$ case. Such contributions only appear at the one-loop
level, which motivates us to undertake the computation of radiative corrections to the pseudoscalar mass in both models.

\section{One-loop potential and one-loop scalar masses}
\label{sec:onel}

In this paper we will closely follow the formalism developed by Stephen P. Martin in a series of
papers~\cite{Martin:1997ns,Martin:2003it,Martin:2003qz}. In this section we will briefly review the formalism required for our purposes
and refer the reader to those references for further details. The one-loop effective potential $V_1$ is given by
\beq
V_1 \,=\, V_0 + \Delta V_1 \,,
\eeq
where we have already written the tree-level potential $V_0$ in eq.~\eqref{eq:potU1} and the one-loop contribution is
\beq
\Delta V_1\,=\, \frac{1}{64\pi^2}\,\sum_\alpha\,n_\alpha \,m_\alpha^4 \,\left[{\log}{\left(\frac{m_\alpha^2}{\mu^2}\right)}\,-\,\frac{3}{2}\right]\,,
\label{eq:pot1}
\eeq
where we are working in the Landau gauge and considering a Dimensional Reduction regularization scheme~\footnote{Had we used Dimensional Regularization,
the factors of 3/2 in $\Delta V_1$ would be replaced by 5/6 for the gauge boson contributions.}. In this expression $\mu$ is an
arbitrary renormalization scale (typically considered of the order of the largest masses of the theory, to numerically minimize the size of the
logarithms present in this expression) and the $\alpha$ index runs through all particles in the model. The $m_\alpha$ are the tree-level masses
for each particle, and the factors $n_\alpha$ are, for a particle of spin $s_\alpha$, given by
\beq
n_\alpha\,=\, (-1)^{2 s_\alpha} \,(2s_\alpha + 1) \, C_\alpha\,Q_\alpha\,,
\label{eq:enes}
\eeq
with $C_\alpha = 3$ for particles with colour and $1$ for all others; and $Q_\alpha = 2$ for particles with electric charge and $1$ for neutral ones. The
one-loop minimization equations are therefore
\beq
\frac{\partial V_1}{\partial v_i}\,=\,\frac{\partial V_0}{\partial v_i}\,+\, \frac{\partial \Delta V_1}{\partial v_i}\,=\,0
\eeq
and a simple calculation gives
\bea
\left(m^2_{11} \,+\, \frac{1}{2} \lambda_1 v_1^2 + \frac{1}{2} \lambda_{34}  v_2^2 \right) v_1 \,+\,
\frac{1}{32\pi^2}\,\sum_\alpha\,n_\alpha \,m_\alpha^2 \,\frac{\partial m^2_\alpha}{\partial v_1}\,\left[{\log}{\left(\frac{m_\alpha^2}{\mu^2}\right)}\,-\,1\right]
& = & 0\,,
\nonumber \\
\left(m^2_{22} \,+\, \frac{1}{2} \lambda_2 v_2^2 + \frac{1}{2} \lambda_{34} v_1^2 \right) v_2 \,+\,
\frac{1}{32\pi^2}\,\sum_\alpha\,n_\alpha \,m_\alpha^2 \,\frac{\partial  m^2_\alpha}{\partial v_2}\,\left[{\log}{\left(\frac{m_\alpha^2}{\mu^2}\right)}\,-\,1\right]
& = & 0\,.
\label{eq:min1}
\eea
Numerically, these equations may be solved for the values of $m^2_{11}$ and $m^2_{22}$ once all other couplings and vevs are chosen~\footnote{Notice, however, that
those two parameters enter into the definitions of the squared scalar masses present in the one-loop terms above.}.
The tree-level masses depend on the vevs, so their derivatives can be obtained easily from eqs.~\eqref{eq:charg}, \eqref{eq:cpeven}, \eqref{eq:pseud}
and~\eqref{eq:gauge} for the scalars and gauge bosons. In section~\ref{sec:onelf} we will show how to handle the fermionic masses. In all of these expressions,
the tree-level expressions for each particle mass are evaluated with vevs which satisfy the one-loop minimization equations. As mentioned earlier, the tree-level
masses of eqs.~\eqref{eq:charg},~\eqref{eq:cpeven} and~\eqref{eq:pseud} evaluated at the tree-level minimum given by eqs.~\eqref{eq:min0} yield
three massless Goldstone bosons and a massless pseudoscalar -- the former due to spontaneous electroweak symmetry breaking, the latter from the spontaneous breaking
of the $U(1)$ global symmetry. In the evaluation of the one-loop effective potential, or its derivatives, tree-level masses are evaluated at the one-loop
minimum and therefore no massless eigenvalues are expected. But when the one-loop scalar mass matrix is evaluated at the one-loop minimum we expect to again find
three massless eigenvalues corresponding to the Goldstone bosons, and again a massless pseudoscalar from spontaneous breaking of $U(1)$.

In~\cite{Martin:2003it,Martin:2003qz} a method of computing the one-loop scalar self-energies was presented. The procedure uses a mass-independent
renormalization scheme, in which observables -- such as masses
or cross sections -- are outputs, but the inputs of the model are renormalized running couplings and mass parameters. This method is ideal
for situations where tree-level predictions for the mass of a given particle are expected to be substantially altered through
radiative corrections, albeit in a perturbative manner. For instance, in SUSY models the Higgs mass is predicted to be less than
the mass of the $Z$ boson at tree-level -- but one-loop contributions to that mass can substantially enhance it, and make
the SUSY value compatible with the current experimental value~\cite{Ellis:1990nz,Ellis:1991zd}. Likewise, this method is appropriate
for the question we are considering in this paper -- to investigate whether the tree-level prediction, $m_A = 0$, is preserved by loop
corrections both for the case of a $U(1)$-invariant lagrangian and a $Z_3$-invariant one.

The physical masses are independent
of both gauge and external momentum, and are defined as the poles of the propagator of the particle in question.
Following~\cite{Martin:2003it,Martin:2003qz},
for a theory with $n$ scalars with tree-level masses $m(0)_i$, the one-loop mass $m(1)_k$ is the value of $s$ which solves
\begin{eqnarray}
{\mbox{Det}}{\left[ \left(m(0)_i^2- s\right)
\delta_{ij} +  \frac{1}{16 \pi^2}\, {\Pi_{ij}}{\left(m(0)_k^2\right)} \right]} = 0\,,
\label{eq:onelmass}
\end{eqnarray}
where $s = m(1)^2_k = -p^2$, $p$ being the external momentum, and $\Pi_{ij}$ are the scalar self-energies. Since the self-energies
$\Pi_{ij}$ have a complicated dependence on the variable $s$, this equation will typically be solved in an iterative way: start with $s = 0$
and compute the eigenvalues of the one-loop mass matrices, take those for the new values of $s$ and repeat the process until $s$
converges to a single value for each mass. This formalism is valid
for a general theory containing an arbitrary number of scalars (S) $R_i$, spin-1/2 Weyl fermions (F) $\psi_I$ and gauge bosons (G)
$A_a^\mu$~\footnote{These fields correspond to the eigenstates of the tree-level mass matrices for scalars, fermions and gauge bosons,
and each will be described by, respectively, lowercase indices $i, j, k, l$, uppercase indices $I, J, K, L$ and lowercase indices $a, b$.},
interacting according to the partial Lagrangians
\begin{eqnarray}
{\cal L}_{\rm S} &=& - \frac{1}{6} \lambda^{ijk} R_i R_j R_k
- \frac{1}{24} \lambda^{ijkl} R_i R_j R_k R_l\,,  \label{eq:LS}
\\
{\cal L}_{\rm SF} &=& - \frac{1}{2} y^{IJk} \psi_I \psi_J R_k + {\rm
h.c.}
\,, \label{eq:LF}
\\
{\cal L}_{\rm SG} &=&
- g^{aij} A^\mu_a R_i \partial_\mu R_j
-\frac{1}{4} g^{abij}  A^\mu_a A_{b\mu} R_i R_j
-\frac{1}{2} g^{abi} A^\mu_a A_{b\mu} R_i
\,.
\label{eq:LSG}
\end{eqnarray}
The conventions are clear: ${\cal L}_{\rm S}$ describes the self interactions between the scalars $R_i$, and the symmetric tensors $\lambda^{ijk}$
and $\lambda^{ijkl}$ will be deduced from the scalar potential of eq.~\eqref{eq:potU1} for the $U(1)$ and $Z_3$ models. ${\cal L}_{\rm SF}$
describes the interactions between scalars and fermions and is therefore a different way of expressing the Yukawa lagrangian of
eq.~\eqref{eq:yuk}, and the $y^{IJk}$ couplings will be related to the $\Gamma$ and $\Delta$ matrices of~\eqref{eq:yuk}.
The $y^{IJk}$ tensors are symmetric on the indices $I$, $J$, and raising/lowering their indices is tantamount to complex conjugation,
{\em i.e} $y^{IJk} = y_{IJk}^*$.

Interactions
between gauge bosons and scalars are contained in ${\cal L}_{\rm SG}$, and the $g^{abi}$ and $g^{abij}$ couplings will be deduced from
the scalar kinetic terms of eq.~\eqref{eq:kin}. With these definitions the one-loop scalar self-energies are found to be~\cite{Martin:2003it,Martin:2003qz}
\begin{eqnarray}
\Pi_{ij}
&=&
\frac{1}{2} \lambda^{ijkk} A (m_k^2) - \frac{1}{2} \lambda^{ikl} \lambda^{jkl} B(m_k^2,m_l^2)
\nonumber \\ &&
+ g^{aik} g^{ajk} B_{SG} (m_k^2, m_a^2)
+ \frac{1}{2} g^{aaij} A_{G} (m_a^2)
+ \frac{1}{2} g^{abi} g^{abj} B_{GG} (m_a^2, m_b^2)
\nonumber \\ &&
+ {\rm Re} \left[ y^{KLi} y_{KLj} \right] B_{FF} (m_K^2, m_L^2)
+ {\rm Re}  \left[ y^{KLi} y^{K'L'j} M_{KK'} M_{LL'}\right]
   B_{\bar{F}\bar{F}} (m_K^2, m_L^2)\,,
\label{eq:Piij}
\end{eqnarray}
where the $A$, $B$ are functions of $s$ and masses, and their explicit expressions are shown in appendix~\ref{ap:def}.
The first line of the equation above includes all scalar-only contributions to the
self-energies, $\Pi^S_{ij}$; the second line all contributions involving gauge bosons, $\Pi^G_{ij}$; and all fermionic
contributions are gathered in the third line, $\Pi^F_{ij}$, which also involves $M_{IJ}$, off-diagonal elements figuring in fermionic propagators in
the Weyl formalism (see~\cite{Martin:1997ns,Martin:2003it} for further details).
Since in this work we are interested in the loop corrections to
the pseudoscalar mass, we use the tree-level mass matrix from eq.~\eqref{eq:pseud} and obtain, for the one-loop mass matrix evaluated
at the desired minimum, the following expression
\bea
\left[M^2_{A} \right] & = & \quad \begin{pmatrix}
m^2_{11} \,+\,\tfrac{1}{2} \lambda_1 v_1^2 + \frac{1}{2} \lambda_{34} v_2^2 \,+\, \tfrac{1}{16\pi^2}\,\Pi_{G^0G^0} &  \tfrac{1}{16\pi^2}\,\Pi_{G^0A} \\[8pt]
\tfrac{1}{16\pi^2}\,\Pi_{A G^0} & m^2_{22} \,+\, \frac{1}{2} \lambda_2 v_2^2 + \frac{1}{2} \lambda_{34} v_1^2 \,+\,\tfrac{1}{16\pi^2}\,\Pi_{AA}
\end{pmatrix} \nonumber \\
 & = & \frac{1}{16\pi^2} \begin{pmatrix}
 X_{G^0G^0} &   \Pi_{G^0A} \\[8pt]
 \Pi_{A G^0} & X_{AA}
\end{pmatrix}
\label{eq:mA1l}
\eea
with
\bea
X_{G^0G^0} &=& \Pi_{G^0G^0} \,-\, \displaystyle{\sum_\alpha}\, n_\alpha \frac{m_\alpha^2}{2 v_1} \,\frac{\partial m^2_\alpha}{\partial
v_1}\,\left[{\log}{\left(\frac{m_\alpha^2}{\mu^2}\right)}\,-\,1\right]\,, \nonumber \\
X_{AA} &=& \,\,\Pi_{AA} \,-\, \displaystyle{\sum_\alpha}\, n_\alpha \,\frac{m_\alpha^2}{2 v_2} \,\frac{\partial  m^2_\alpha}{\partial
v_2}\,\left[{\log}{\left(\frac{m_\alpha^2}{\mu^2}\right)}\,-\,1\right]\,,
\label{eq:Xizes}
\eea
where we have already used the one-loop minimization conditions from eq.~\eqref{eq:min1} and made the choice $G^0 \equiv \varphi_7$
and $A \equiv \varphi_8$ (see eq.~\eqref{eq:doub}; this choice is simplified by the fact that the tree-level pseudoscalar mass
matrix~\eqref{eq:pseud} is diagonal).

We now present detailed calculations of each of the contributions to the loop corrections -- scalar, gauge and fermionic -- with the
expectation that:
\begin{itemize}
\item Since the $U(1)$ and $Z_3$ models are identical in all regards in the scalar and gauge sectors, the respective contributions to
$[M^2_{A}]$ must keep the pseudoscalar massless.
\item Using $U(1)$-symmetric Yukawa matrices should yield fermionic contributions to the self-energies such
that both the neutral Goldstone and the pseudoscalar remain massless. From Goldstone's theorem a massless pseudoscalar is to be expected
since a continuous  symmetry of the lagrangian has been spontaneously broken.
\item If $Z_3$-symmetric Yukawa matrices are considered then the fermionic contributions to the self-energies should be such that while
the neutral Goldstone remains massless the pseudoscalar mass receives a finite, non-zero contribution. This will be a verification of
Goldstone's theorem  since the lagrangian symmetry being spontaneously broken is discrete.
\end{itemize}
One important detail to consider: since we are interested in loop corrections to the pseudoscalar mass matrix, which includes a massless
Goldstone boson, the calculation will in principle yield two different values of $s$.

\section{One-loop masses: scalar and gauge sectors}
\label{sec:onelsg}

We will first deal with the contributions to the one-loop minimization conditions of eq.~\eqref{eq:min1}. These are obvious for the gauge
bosons: their masses are given in eqs.~\eqref{eq:gauge}, so their derivatives are trivial to obtain and yield
\beq
\frac{\partial \Delta V_1^{G}}{\partial v_i}\,=\,
\frac{1}{32\pi^2}\,\frac{v_i}{v^2}\,\left[ 6\,m_Z^2 \,A(m_Z^2) \,+\,
12 \,m_W^2  \,A(m_W^2)\right]
\label{eq:minG}
\eeq
with all spin and charge degrees of freedom already taken into account and the $A(x)$ function is defined as
\beq
A(x) \,=\,x\,\left[{\log}{\left(\frac{x}{\mu^2}\right)}\,-\,1\right]\,.
\eeq
The scalar contributions are equally simple for the neutral Goldstone
and pseudoscalar -- using eqs.~\eqref{eq:pseud} we obtain
\bea
\frac{\partial m^2_{G^0}}{\partial v_1} \,=\, \lambda_1 \, v_1 &,&
\frac{\partial m^2_{G^0}}{\partial v_2} \,=\, \lambda_{34} \, v_2 \nonumber \\
\frac{\partial m^2_{A}}{\partial v_1} \,=\, \lambda_{34}  \, v_1 &,&
\frac{\partial m^2_{A}}{\partial v_2} \,=\, \lambda_2 \, v_2\, .
\eea
As for the contributions from the charged and CP-evens scalars, we see from eqs.~\eqref{eq:charg} and~\eqref{eq:cpeven} that
their squared masses are the eigenvalues of real symmetric $2\times 2$ matrices of the form
\beq
\left[M^2_S \right]\,=\,\begin{pmatrix}
a &   b\\
b & c
\end{pmatrix}
\label{eq:mat}
\eeq
with vev-dependent coefficients $a$, $b$ and $c$. The corresponding eigenvalues are therefore
the solutions of the characteristic equation
\beq
F(x,v_i)\,=\, x^2 \,-\,(a + c)\,x \,+\, ac - b^2\,=\,0
\label{eq:charac}
\eeq
which gives us
\beq
m^2_{1,2} \,=\,\frac{1}{2}\,\left( a\,+\,c \,\pm \,\sqrt{(a - c)^2 + 4 b^2}\right)\,.
\eeq
The lowest of the eigenvalues (associated with the ``$-$" sign in this equation) correspond to the lightest CP-even
scalar $h$ and to the tree-level charged Goldstone $G^\pm$ squared mass.
Analytical expressions for the derivatives of these eigenvalues may be obtained directly from the expression above or, applying the implicit
function theorem to eq.~\eqref{eq:charac}, from
\beq
\frac{\partial m^2_{1,2}}{\partial v_i}\,=\,\frac{(a_i + c_i) m^2_{1,2} \,+\, 2 b b_i - c a_i - a c_i }{m^2_{1,2} - m^2_{2,1}}
\eeq
where $a_i$, $b_i$ and $c_i$ represent the derivatives of the respective coefficients with respect to $v_i$, and they
are very simple expressions of the $\lambda$ couplings and vevs. This expression for the derivatives of the squared masses is
quite useful if one wants to perform analytical checks but it is also helpful when performing numerical calculations.
Putting everything together we have
\bea
(32\pi^2)\frac{\partial \Delta V_1^{S}}{\partial v_1} &=&
\left[ 2\,\frac{(\lambda_1 + \lambda_3) v_1  m_{H^\pm}^2 + \lambda_4 v_2 b_\pm - (\lambda_1 c_\pm + \lambda_3 a_\pm) v_1 }{
m_{H^\pm}^2 - m_{G^\pm}^2}\,A(m^2_{H^\pm}) \right. \nonumber \\
 & & +\,
2\,\frac{(\lambda_1 + \lambda_3) v_1  m_{G^\pm}^2 + \lambda_4 v_2 b_\pm - (\lambda_1 c_\pm + \lambda_3 a_\pm) v_1  }{m_{G^\pm}^2 - m_{H^\pm}^2} \,A(m^2_{G^\pm})
\nonumber \\
 & & +\,
\frac{(3 \lambda_1 + \lambda_{34}) v_1 m_{h}^2 + 2 \lambda_{34} v_2 b_H - (3 \lambda_1 c_H
 + \lambda_{34} a_H) v_1}{m_{h}^2 - m_{H}^2}\,A(m^2_{h})  \nonumber \\
 & & +\,
\frac{(3 \lambda_1 + \lambda_{34}) v_1 m_{H}^2 + 2 \lambda_{34} v_2 b_H - (3 \lambda_1 c_H
 + \lambda_{34} a_H) v_1}{m_{H}^2 - m_{h}^2} \,A(m^2_{H})
 \nonumber \\
 & & +\,
 \left. \frac{}{} \lambda_1 v_1 \,A(m^2_{G^0})\,+\,\lambda_{34} v_1 \, A(m^2_A)
\right]\,,
\label{eq:der1Lv1}\\
(32\pi^2)\frac{\partial \Delta V_1^{S}}{\partial v_2} &=&
\left[ 2\,\frac{(\lambda_2 + \lambda_3) v_2  m_{H^\pm}^2 + \lambda_4 v_1 b_\pm - (\lambda_3 c_\pm + \lambda_2 a_\pm) v_2}{
m_{H^\pm}^2 - m_{G^\pm}^2}\,A(m^2_{H^\pm}) \right. \nonumber \\
 & & +\,
2\,\frac{(\lambda_2 + \lambda_3) v_2  m_{G^\pm}^2 + \lambda_4 v_1 b_\pm - (\lambda_3 c_\pm + \lambda_2 a_\pm) v_2}{m_{G^\pm}^2 - m_{H^\pm}^2} \,A(m^2_{G^\pm})
\nonumber \\
 & & +\,
\frac{(3 \lambda_2 + \lambda_{34}) v_2 m_{h}^2 + 2 \lambda_{34} v_1 b_H - (\lambda_{34} c_H
 + 3 \lambda_2 a_H) v_2}{m_{h}^2 - m_{H}^2}\,A(m^2_{h})  \nonumber \\
 & & +\,
\frac{(3 \lambda_2 + \lambda_{34}) v_2 m_{H}^2 + 2 \lambda_{34} v_1 b_H - (\lambda_{34} c_H
 + 3 \lambda_2 a_H) v_2}{m_{H}^2 - m_{h}^2} \,A(m^2_{H})
 \nonumber \\
 & & +\,
 \left. \frac{}{} \lambda_{34} v_2 \, A(m^2_{G^0})\,+\,\lambda_2 v_2 \, A(m^2_A)
\right]\,,
\label{eq:der1Lv2}
\eea
where $\{a_\pm , b_\pm , c_\pm\}$ ($\{a_H , b_H , c_H\}$) are the entries of
matrix~\eqref{eq:charg} (\eqref{eq:cpeven}) following the convention of eq.~\eqref{eq:mat}.

The self-energies require that we obtain the tensors $\lambda^{ijk}$, $\lambda^{ijkl}$, $g^{aij}$,
$g^{abi}$ and $g^{abij}$. To do so we must express the tree-level potential~\eqref{eq:potU1} and kinetic terms~\eqref{eq:kin}
in terms of mass eigenstates, both of scalars and gauge bosons. As we already discussed, the neutral Goldstone and pseudoscalar
are identified with the scalar components $\varphi_7$ and $\varphi_8$. The charged component fields $\varphi_{1 ... 4}$ give
rise to two charged scalars and Goldstone bosons through the diagonalization angle $\beta$ of
matrix~\eqref{eq:charg}~\footnote{Readers will notice that though we use the notation $H^\pm$, $G^\pm$, these
are real fields, not the complex ones usually used to represent charged scalars. The formalism
of~\cite{Martin:1997ns,Martin:2003it,Martin:2003qz}, however, requires that we indeed use real fields for
all eigenstates.},
\bea
H^+ \,=\, c_\beta \varphi_1 - s_\beta \varphi_2 &,&  G^+ \,=\,s_\beta \varphi_1 + c_\beta \varphi_2 \nonumber \\
H^- \,=\, c_\beta \varphi_3 - s_\beta \varphi_4 &,&  G^- \,=\,s_\beta \varphi_3 + c_\beta \varphi_4
\label{eq:beta}
\eea
where we use the notation $c_x = \cos x$ and $s_x = \sin x$. Since we are considering a one-loop
minimization of the potential the angle $\beta$ is {\em not} the one usually considered in 2HDM
calculations, meaning that here $\tan\beta \neq v_2/v_1$. The two CP-even tree-level eigenstates
are likewise defined through the diagonalization angle $\alpha$ of~\eqref{eq:cpeven},
\beq
h\,=\, c_\alpha \varphi_5 - s_\alpha \varphi_6 \quad \,,\, \quad H\,=\, s_\alpha \varphi_5 +
c_\alpha \varphi_6\,.
\eeq
With these definitions and the conventions of eq.~\eqref{eq:LS}--\eqref{eq:LSG}, the tensors required
are simply the derivatives of the lagrangian with respect to the tree-level mass eigenstates, with those fields set to zero.
So for instance,
\beq
\lambda^{AAhh}\,=\,\frac{\partial^4 V_0}{\partial^2 A \partial^2 h} \,=\,
s^2_\alpha\,\lambda_2\,+\, c^2_\alpha\,  \lambda_{34}
\eeq
and
\beq
\lambda^{AG^+ H^-}\,=\,\frac{\partial^3 V_0}{\partial A \partial G^+ \partial H^-} \,=\,
-\,\frac{\lambda_4\,v_1}{2}\,.
\eeq
All non-zero coefficients $\lambda$ relevant for the current calculation are listed in Appendix~\ref{ap:def}. We now show
the final expressions for the scalars' contributions to the self-energies, leaving intermediate calculations to the reader.
For the off-diagonal term, we have
\beq
\Pi^S_{AG^0}\,=\,\Pi^S_{G^0A} \,=\, \frac{\lambda_4}{4} \left\{ 4 s_\beta c_\beta
\left[  A(m^2_{G^\pm}) - A(m^2_{H^\pm}) \right] \,+\,
\lambda_4 \,v_1 v_2 \left[ B(m^2_{H^\pm}, m^2_{G^\pm}) \,+\, B(m^2_{G^\pm}, m^2_{H^\pm}) \right]\right\}\,,
\eeq
where the $B(x,y)$ function is defined in eq.~\eqref{eq:B}.
We can now use the definition of the diagonalization angle $\beta$, eq.~\eqref{eq:beta}, of the charged mass matrix,
eq.~\eqref{eq:charg}, to obtain the following relation:
\beq
\lambda_4\,=\, \frac{2 s_\beta c_\beta}{v_1 v_2} \,\left(m_{G^\pm}^2 - m_{H^\pm}^2\right)\,.
\label{eq:l41}
\eeq
For $s = 0$~\footnote{We should definitely perform this computation with $s = 0$ to obtain the mass of the neutral Goldstone boson at
one-loop. As we will see, however, when only the scalar and gauge contributions to the self-energies are considered, the choice $s = 0$ yields
two zero eigenvalues, which proves that the pseudoscalar is also found to be massless.} the $B$ function is simply $B(x,y) = (A(x) - A(y))/(y - x)$
and therefore it follows that
\beq
\Pi^S_{AG^0}\,=\, \lambda_4\,s_\beta c_\beta\,\left[ A(m^2_{G^\pm}) - A(m^2_{H^\pm})\,+\, \left(m_{G^\pm}^2 - m_{H^\pm}^2\right)\,
\frac{A(m^2_{H^\pm}) - A(m^2_{G^\pm})}{m_{G^\pm}^2 - m_{H^\pm}^2} \right]\,=\, 0\,.
\eeq
For the diagonal terms we obtain
\bea
\Pi^S_{G^0 G^0} & = &
\frac{1}{2} \left[\frac{}{} \lambda_{34} A(m^2_A) + 3 \lambda_{1} A(m^2_{G^0})
    + \left( \lambda_{34} c^2_\alpha + \lambda_{1} s^2_\alpha \right) A(m^2_{H})
    + \left( \lambda_{1} c^2_\alpha + \lambda_{34} s^2_\alpha \right) A(m^2_{h})
    \right] \nonumber \\
  & & +\,\left( \lambda_{3} c^2_\beta + \lambda_{1} s^2_\beta \right) A(m^2_{G^\pm})
    + \left( \lambda_{1} c^2_\beta + \lambda_{3} s^2_\beta \right) A(m^2_{H^\pm}) - \frac{v_2^2 \lambda_4^2}{2}\, B(m^2_{G^\pm},m^2_{H^\pm}) \nonumber \\
  & & - \,\left( \lambda_{1} v_1 s_\alpha +  \lambda_{34} v_2 c_\alpha \right)^2 B(m^2_{G^0},m^2_{H}) - \left(  \lambda_{1} v_1 c_\alpha -  \lambda_{34} v_2 s_\alpha \right)^2 B(m^2_{G^0},m^2_{h})\,,
  \label{eq:PIG0G0}\\
\Pi^S_{AA} & = &
\frac{1}{2} \left[ \frac{}{}  3 \lambda_{2} A(m^2_A) +  \lambda_{34} A(m^2_{G^0})
    + \left( \lambda_{2} c^2_\alpha + \lambda_{34} s^2_\alpha \right) A(m^2_{H})
    + \left( \lambda_{34} c^2_\alpha + \lambda_{2} s^2_\alpha \right) A(m^2_{h})
    \right] \nonumber \\
  &  & + \,\left( \lambda_{2} c^2_\beta + \lambda_{3} s^2_\beta \right) A(m^2_{G^\pm})
    + \left( \lambda_{3} c^2_\beta + \lambda_{2} s^2_\beta \right) A(m^2_{H^\pm}) - \frac{ v_1^2 \lambda_4^2}{2}\, B(m^2_{G^\pm},m_{H^\pm}) \nonumber \\
  &  & - \,\left( \lambda_{34} v_1 s_\alpha + \lambda_{2} v_2 c_\alpha \right)^2 B(m^2_{A},m^2_{H}) - \left( \lambda_{34} v_1 c_\alpha - \lambda_{2} v_2 s_\alpha \right)^2 B(m^2_{A},m^2_{h})\,.
\label{eq:PIAA}
\eea
These expressions must now be plugged into eq.~\eqref{eq:mA1l} and~\eqref{eq:Xizes} along with the explicit one-loop potential derivatives
of eq.~\eqref{eq:der1Lv1} and~\eqref{eq:der1Lv2}. With $s = 0$, and expressing all $B$ functions in terms of $A$ functions, a careful calculation will see
a series of cancelations occur -- these arise from the fact that the diagonalization angles $\beta$ and $\alpha$ are not independent
of the masses, vevs and couplings of the model, as we saw in eq.~\eqref{eq:l41}. The following relations were found to be useful
for this calculation:
\bea
\lambda_1 &=& \frac{1}{v_1^2}\,\left( c_\alpha^2 m^2_h \,+\, s_\alpha^2 m^2_H \,-\, m^2_{G^0}\right)\,, \nonumber \\
\lambda_2 &=& \frac{1}{v_2^2}\,\left( s_\alpha^2 m^2_h \,+\, c_\alpha^2 m^2_H \,-\, m^2_A\right)\,, \nonumber \\
\lambda_{34} &=& \frac{m^2_H - m^2_h}{v_1 v_2}\,s_\alpha c_\alpha\,.
\label{eq:lambs}
\eea
The end result is that all entries of the $2\times 2$ matrix of~\eqref{eq:mA1l} become equal to {\em zero}. The readers are invited to verify
this analytic result in full. To illustrate the calculation we show explicitly the vanishing of terms involving the contributions 
from $A$ and $G^0$
in $\Pi_{G^0 G^0}$ -- keeping only those terms in eqs.~\eqref{eq:Xizes},~\eqref{eq:der1Lv1} and~\eqref{eq:PIG0G0}, we obtain
\bea
X_{G^0 G^0} &=& \frac{1}{2} \left[ \lambda_{34} A(m^2_A) + 3 \lambda_{1} A(m^2_{G^0})\right]\,- \,\left(
\lambda_{1} v_1 s_\alpha +  \lambda_{34} v_2 c_\alpha  \right)^2
B(m^2_{G^0},m^2_{H}) \nonumber \\
 & & - \left( \lambda_{1} v_1 c_\alpha - \lambda_{34} v_2 s_\alpha \right)^2 B(m^2_{G^0},m^2_{h})\,-\,
 \frac{1}{2 v_1} \left\{ \lambda_1 v_1 \,A(m^2_{G^0})\,+\,\lambda_{34} v_1 \, A(m^2_A) \right\} \,,
\eea
where the term in curly brackets comes from the derivatives of the one-loop potential. We see that the terms in $A(m_A^2)$ cancel out and,
using the relations~\eqref{eq:lambs}, the factors multiplying the $B$ functions above are given by
\bea
\lambda_{1} v_1 s_\alpha + \lambda_{34} v_2 c_\alpha  & = &
\frac{m^2_H - m^2_{G^0}}{v_1}\,s_\alpha \,, \nonumber \\
\lambda_{1} v_1 c_\alpha -  \lambda_{34} v_2 s_\alpha  & = &
\frac{m^2_h - m^2_{G^0}}{v_1}\,c_\alpha\,.
\eea
Thus,  using the explicit $s = 0$ form of the $B$ functions and keeping only the $A(m_{G^0}^2)$ terms, this becomes
%
%
\bea
X_{G^0 G^0} &=& \lambda_{1} A(m^2_{G^0})\,- \,\frac{s^2_\alpha}{v^2_1} \left(m^2_H - m^2_{G^0}\right)^2\,
\frac{A(m^2_{G^0})}{m_H^2 - m^2_{G^0}} -
\frac{c^2_\alpha}{v^2_1} \left(m^2_h - m^2_{G^0}\right)^2\, \frac{A(m^2_{G^0}) }{m_h^2 - m^2_{G^0}} \nonumber \\
& = & \lambda_{1} A(m^2_{G^0}) \,- \, \frac{1}{v_1^2} \left( c_\alpha^2 m^2_h \,+\, s_\alpha^2 m^2_H \,-\, m^2_{G^0}\right)\, A(m^2_{G^0})\,=\, 0\,,
\eea
where in the final step we once again used the first of eqs.~\eqref{eq:lambs}. Similar calculations are performed for the rest of the terms
and show that all one-loop scalar contributions to the pseudoscalar mass matrix vanish at the minimum.

The same in fact holds for the contributions stemming from the gauge sector.
The $g$ tensors detailing the scalar and gauge interactions are, as before, obtained through derivatives
of the kinetic terms with respect to scalar and gauge mass eigenstate fields, setting those fields to zero. For instance,
\beq
g^{AAW^+W^-}\,=\,-\,\frac{\partial^4 {\cal L}_K}{\partial A^2 \partial W^+ \partial W^-} \,=\,\frac{g^2}{2}\,=\,\frac{2 m^2_W}{v^2}\,.
\eeq
We also obtain $g^{abA} = g^{abG^0} = 0$, for all gauge bosons $a$, $b$, and another example is
\beq
g^{ZG^0 h}\,=\,-\,\frac{\partial^3 {\cal L}_K}{\partial Z_\mu \partial G^0 \partial (\partial^\mu h)}
\,=\,-\, \frac{m_Z}{v}\,c_\alpha\,.
\eeq
All non-zero coefficients $g$ relevant for the current calculation are listed in Appendix~\ref{ap:def}. With these coefficients we
can proceed to compute the contributions of the gauge sector to the self-energies in eq.~\eqref{eq:Piij}. For the off-diagonal
elements of the pseudoscalar mass matrix we obtain
\beq
\Pi^G_{AG^0} \, = \,\frac{m^2_Z  }{v^2} s_\alpha c_\alpha \left[ B_{SG}(m_H^2, m_Z^2) - B_{SG}(m_h^2, m_Z^2) \right]\,,
\eeq
where the $B_{SG}$ function is defined in eq.~\eqref{eq:BSV}. It is then easy to see that in the Landau gauge and for the limit of zero external
momentum, $s = 0$, one obtains $B_{SG}(x,y) = 0$, and therefore $\Pi^{G}_{AG^0} = 0$.
Once again we find that the one-loop contribution to the off-diagonal element of the pseudoscalar mass matrix is zero. As for the
diagonal elements, we obtain
\bea
\Pi^G_{AA} & = & \frac{m^2_Z}{v^2} \left[ s^2_\alpha B_{SG}(m_{h}^2, m_{Z}^2)
    + c_\alpha^2  B_{SG}(m_H^2,m_Z^2)
    + A_G(m^2_Z) + 2 c_W^2 A_G(m^2_W) \right] \,, \\
\Pi^G_{G^0 G^0} & = &\frac{m^2_Z}{v^2} \left[ c^2_\alpha B_{SG}(m_{h}^2, m_{Z}^2)
    + s_\alpha^2  B_{SG}(m_H^2,m_Z^2)
    + A_G(m^2_Z) + 2 c_W^2 A_G(m^2_W) \right] \,,
\eea
where $c_W = m_W/m_Z$. As mentioned before, the $B_{SG}$ function is zero for $s = 0$ and in the Landau gauge, and the function
$A_G(x)$ is equal (in the same circunstances and for a DRED renormalization scheme) to $3 A(x)$
(see eq.~\eqref{eq:AG})~\footnote{This factor of ``3" counts the spin
degrees of freedom of the gauge bosons, see eq.~\eqref{eq:enes}.}. Thus we obtain
\beq
\Pi^G_{AA} \,\,=\,\, \Pi^G_{G^0 G^0} \,\,=\,\,  \frac{3 m^2_Z}{v^2} \left[ A(m^2_Z) + 2 c_W^2 A(m^2_W) \right]\,.
\eeq
Once this expression is inserted into eq.~\eqref{eq:Xizes} along with the contributions to the minimization conditions involving gauge bosons, from eq.~\eqref{eq:minG}, all terms in $A(x)$ cancel out.

Thus we conclude that, when considering the one-loop contributions from scalars and gauge bosons to the pseudoscalar mass matrix,
said matrix remains identically equal to zero, as it was at tree-level. Therefore we obtain a massless neutral Goldstone boson,
as required by electroweak gauge symmetry breaking, and a massless pseudoscalar -- as is to be expected since the lagrangian of the model
has a continuous $U(1)$ symmetry, the spontaneous breaking of which must needs produce a massless particle. And since if the scalar and gauge
sectors the imposition of a $U(1)$ or $Z_3$ symmetries is indistinguishable, these results confirm the validity of Goldstone's theorem --
the imposition of a discrete $Z_3$ symmetry in a scalar + gauge 2HDM leads to a theory with an accidental continuous symmetry, therefore
a massless pseudoscalar follows after spontaneous symmetry breaking. This now changes when we consider, finally, the contributions from
fermions to the self-energies.

\section{One-loop masses: fermion sector}
\label{sec:onelf}

The fermion sector calculations are made more difficult by the fact that we wish to study generic textures of Yukawa matrices. Models
with flavour conservation, such as 2HDMs of Type I (or II, or X, or Y), allow, as we shall see, for a full analytic calculation of one-loop
pseudoscalar mass corrections. But in generic models where FCNCs are present that will not be possible. In fact, since there are three
generations of quarks, the eigenvalues of the quark mass matrices cannot in general be obtained
analytically, unlike the scalar and gauge cases. Our investigation of the fermion sector will therefore perforce be (mostly) numerical -- we will
choose random values for both vevs $v_1$ and $v_2$, obviously respecting $v_1^2 + v_2^2 = v^2$, and random values for the entries of the Yukawa matrices
so that six massive quarks are obtained; we will then compute the value of the one-loop contributions to the self-energies and minimization conditions
and determine numerically whether the pseudoscalar acquires a mass or not.

There is however a semi-analytical way to proceed in dealing with the minimization conditions.
Considering the definitions of the quark mass matrices, eq.~\eqref{eq:Mnp}, and their diagonalization matrices, eq.~\eqref{eq:Mdu}, we see that the
matrices
\beq
H_d\,=\,M_n\,M_n^\dagger\;\;\; , \;\;\; H_u\,=\,M_p\,M_p^\dagger
\eeq
are diagonalized by the unitary left matrices of eq.~\eqref{eq:Mdu}, and their eigenvalues are the squared down and up quark masses,
respectively -- since these matrices are not bi-diagonalized as the matrices $M_n$, $M_p$ are, their eigenvalues are guaranteed to be
real and positive, and may be obtained numerically in a trivial manner.
Notice that for each chosen texture for the matrices $\Gamma_1$ and $\Gamma_2$ ($\Delta_1$ and $\Delta_2$) we can write simple
analytical expressions for each entry of $H_d$ ($H_u$). For the up sector, the squared masses are the solutions $m^2$ of a
cubic characteristic equation of the form
\beq
{\rm{Det}}{\left(H_u\,-\,m^2\,\mathbb{1}_{3\times 3} \right)} \,=\,0\,.
\eeq
Defining the $3\times 3$ matrix $F_u = H_u\,-\,m^2\,\mathbb{1}_{3\times 3}$, this equation may now be be written as
\beq
\mathcal{F}_u(m^2,v_a)\,=\, \sum_{i,j,k=1}^3\,\epsilon_{ijk}\,F_{1i}\,F_{2j}\,F_{3k}\,=\,0
\eeq
and therefore, using the implicit function theorem, we may write
\beq
\frac{\partial m^2}{\partial v_a} \,=\, -\,\frac{ \displaystyle{\frac{\partial \mathcal{F}_u }{\partial v_a} }}{
\displaystyle{\frac{\partial \mathcal{F}_u }{\partial m^2}} }\,.
\label{eq:derimp}
\eeq
The derivatives of $\mathcal{F}_u$ are given by
\beq
\frac{\partial \mathcal{F}_u}{\partial x} \,=\,  \sum_{i,j,k=1}^3\, \left[ \epsilon_{ijk}\,\frac{\partial F_{1i}}{\partial x}\,F_{2j}\,F_{3k} \,+\,\epsilon_{ijk}\,F_{1i}\,\frac{\partial F_{2j}}{\partial x}\,F_{3k}
\,+\,\epsilon_{ijk}\,F_{1i}\,F_{2j}\,\frac{\partial F_{3k}}{\partial x} \right]\,,
\label{eq:derFu}
\eeq
that is, they are given by the sum of the determinants of the three matrices obtained from $F_u$ by replacing one of their lines
by its derivative with respect to the variable $x$. For a given choice of textures for $\Delta_1$ and $\Delta_2$, all of these determinants are
analytical expressions of the entries of those matrices, the vevs and the mass of each quark. Therefore, once the eigenvalues of
$H_u$ have been obtained numerically, we may use them in eqs.~\eqref{eq:derimp} and~\eqref{eq:derFu} and obtain the numerical values of the derivatives
of the squared up quark masses with respect to the vevs. The procedure is analogous for the down sector. The quark contributions to the one-loop
minimization conditions will therefore be
\beq
\frac{\partial \Delta V_1^{F}}{\partial v_i}\,=\,
- \frac{12}{32\pi^2}\,\sum_{\rm{quarks}} \, \frac{\partial m^2_q}{\partial v_i}\,A(m^2_q)\,,
\label{eq:minF}
\eeq
where the factor ``$-12$" stems from eq.~\eqref{eq:enes}.

The self-energies of eq.~\eqref{eq:Piij} use Yukawa couplings from the scalar-fermion interaction lagrangian, eq.~\eqref{eq:LF}, written in terms
of Weyl spinors in the basis of quark mass eigenstates. Since the 2HDM Yukawa lagrangian we wrote, eq.~\eqref{eq:yuk}, or the textures from eqs.~\eqref{eq:tI}
and~\eqref{eq:Yukz3}, are written in terms of Dirac fermion notation, a ``dictionary" between both notations is needed. First we need to rotate the $\Gamma$
and $\Delta$ matrices to the mass basis -- this is accomplished through the rotation matrices in eq.~\eqref{eq:Mdu}, so that
\beq
\bar{\Gamma}_a \,=\, U_{dL}^\dagger\,\Gamma_a\, U_{dR}\;\;\; , \;\;\;
\bar{\Delta}_a \,=\,  U_{uL}^\dagger\,\Delta_a\, U_{uR}\,,
\label{eq:rotmat}
\eeq
where the bars indicate Yukawa matrices in the quark mass basis. In this paper we are interested in loop corrections to the pseudoscalar sector,
so we can restrict ourselves to the interactions of $G^0$ and $A$ in the Yukawa lagrangian of eq.~\eqref{eq:yuk}. Since these two scalar
eigenstates coincide (at tree-level) with the imaginary neutral components of, respectively, $\Phi_1$ and $\Phi_2$ we obtain, for the three
up-type quarks $u_i$ and three down-type quarks $d_j$, the following lagrangian:
\bea
-{\cal L_Y} &=& \left[-\,\frac{i}{\sqrt{2}} \,\left(\bar{\Gamma}_1^\dagger\right)_{ij}\, \bar{d}_i\, P_L \,d_j\,+\,\frac{i}{\sqrt{2}}\,\left(\bar{\Delta}_1^\dagger\right)_{ij}\, \bar{u}_i\, P_L \,u_j
\,+ \,{\rm h.c.} \right]\,G^0 \nonumber \\
 & & +\, \left[-\,\frac{i}{\sqrt{2}}\,\left(\bar{\Gamma}_2^\dagger\right)_{ij}\, \bar{d}_i\, P_L \,d_j\,+\,\frac{i}{\sqrt{2}}\,\left(\bar{\Delta}_2^\dagger\right)_{ij}\, \bar{u}_i\, P_L \,u_j
\,+ \,{\rm h.c.} \right]\, A \, +  \ldots
\label{eq:LagF}
\eea
where we have introduced the usual left and right projectors $P_L$ and $P_R$. Let us now follow the notation of~\cite{Martin:1997ns} and
recall that any Dirac spinor $\Psi$ may be represented by two Weyl spinors, $\xi$ and $\chi$ -- in this language ``$\xi$" will embody the
left-handed component of the Dirac spinor $\Psi$, and ``$\chi^\dagger$" its right-handed one, so that we will have, for two different fermions,
\beq
\overline{\Psi}_i\,P_L\,\Psi_j \,=\, \chi_i\,\xi_j\;\;\; , \;\;\;  \overline{\Psi}_i\,P_R\,\Psi_j \,=\, \xi_i^\dagger\,\chi^\dagger_j\,.
\eeq
To make the connection with the lagrangian from eq.~\eqref{eq:LF}, then, we introduce 12 Weyl spinors, numbered such that odd/even indices correspond to
$\chi$/$\xi$ fields, to wit
\bea
\psi_1 \equiv \chi_u\;,\; \psi_3 \equiv \chi_c\;,\; \psi_5 \equiv \chi_t\;,\; \psi_7 \equiv \chi_d
\;,\; \psi_9 \equiv \chi_s\;,\; \psi_{11} \equiv \chi_b \, , \nonumber \\
\psi_2 \equiv \xi_u\;,\; \psi_4 \equiv \xi_c\;,\; \psi_6 \equiv \xi_t\;,\; \psi_8 \equiv \xi_d
\;,\; \psi_{10} \equiv \xi_s\;,\; \psi_{12} \equiv \xi_b \, ,
\eea
so that, if for instance one selects the top quark interaction terms with $A$ from eq.~\eqref{eq:LagF}, we get
\bea
-{\cal L_Y} &=& \frac{i}{\sqrt{2}}\,\left[ \left(\bar{\Delta}_2^\dagger\right)_{33}\, \bar{t}\, P_L \,t \,+\,
\left(\bar{\Delta}_2^\dagger\right)_{32}\, \bar{t}\, P_L \,c \,+\,
\left(\bar{\Delta}_2^\dagger\right)_{31}\, \bar{t}\, P_L \,u \,+ \,{\rm h.c.}  \right]\,A \,+\,\ldots \nonumber \\
& = &
\left(y^{56A}\,\psi_5\,\psi_6 \,+\, y^{54A}\,\psi_5\,\psi_4 \,+\, y^{52A}\,\psi_5\,\psi_2 \,+\, {\rm h.c.}\right)  \, A \,+\,\ldots \;.
\label{eq:ident}
\eea
One can then identify each entry of the $y^{IJk}$ tensors~\footnote{The $y^{IJk}$ tensors are symmetric on the indices $I, J$,
due to the fact that the product of two Weyl spinors is also symmetric, {\em i.e.} $\xi \chi = \chi \xi$. See~\cite{Martin:1997ns}
for more details.}, for example, one has
\bea
y^{56A}\,=\, y^{65A} & = & \frac{i}{\sqrt{2}}\,\left(\bar{\Delta}_2^\dagger\right)_{33}\,, \nonumber \\
y^{54A}\,=\, y^{45A} & = & \frac{i}{\sqrt{2}}\,\left(\bar{\Delta}_2^\dagger\right)_{32}\,.
\eea
Notice that most of the entries of the $y^{IJk}$ tensors are zero -- all diagonal terms, $y^{IIk} = 0$, due to the way Weyl mass terms
are written; and since we are not considering charge breaking vacua the up and down quark mass matrices will not mix and therefore
$y^{IJk} = 0$ for $I\in \{1,\ldots,6\}$ and simultaneously $J \in \{7,\ldots,12\}$. Further, the coefficients $M^{IJ}$ used in
eq.~\eqref{eq:Piij} are obtained from the mass terms in the lagrangian through
\beq
M^{IJ}\,=\, -\,\frac{\partial^2 \cal{L} }{\partial \psi_I \partial \psi_J}\,,
\label{eq:MIJ}
\eeq
this derivative being computed with all fields set equal to zero.
We therefore have a semi-analytical algorithm to compute one-loop fermionic corrections to scalar mass matrices:
\begin{itemize}
\item Obtain, from symmetry arguments, the Yukawa textures for the $\Gamma$ and $\Delta$ matrices.
\item Choose vevs $v_1$ and $v_2$ and random values for the Yukawa couplings. Numerically obtain
the eigenvalues of the quark mass matrices and their diagonalization matrices.
\item Compute the quark contributions to the minimization equations using the procedure outlined before
eq.~\eqref{eq:minF}.
\item Rotate the Yukawa matrices to the quark mass eigenbasis, eq.~\eqref{eq:rotmat}.
\item Identify the non-zero entries of the tensors $y^{IJk}$ as per the method described in eq.~\eqref{eq:ident}.
\item With the $y^{IJk}$ and quark masses determined, compute the fermionic contributions to the self-energies, $\Pi^F_{ij}$,
from eq.~\eqref{eq:Piij}.
\end{itemize}
We will now apply this procedure to textures which are obtained from $U(1)$ and $Z_3$ symmetries. To do this, we will
use the work of reference~\cite{Ferreira:2010ir}, where the effect of all 2HDM abelian symmetries on the Yukawa matrices
was obtained.

\subsection{$U(1)$-symmetric Yukawa textures}

With the $U(1)$ symmetry extended to the Yukawa sector, not just the scalar and gauge sectors, we have a
lagrangian invariant under a continuous symmetry. Goldstone's theorem therefore implies that the pseudoscalar must remain
massless at loop level. Let us see that this is indeed the case in an explicit simple example, that of a 2HDM with Type-I
Yukawa couplings inforced by a $U(1)$ symmetry, {\em i.e.} the textures shown in eq.~\eqref{eq:tI}. Unlike the general case
treated above, for Type-I Yukawas a single doublet (in this case $\Phi_2$) couples to fermions. Therefore the symmetry enforces
$\Gamma_1 = \Delta_1 = 0$ and the up (down) Yukawa matrices are directly proportional to the up (down) quark mass matrices --
therefore the quark mass eigenbasis is also the basis where the Yukawa matrices are diagonal, and no FCNC occurs. All quark masses
are therefore of the form
\beq
m_q\,=\,\frac{\lambda_q}{\sqrt{2}}\,v_2\,,
\eeq
where $\lambda_q$ is a diagonal element of $\Gamma_2$/$\Delta_2$ in the quark mass basis.
Substantial simplifications then occur:
\begin{itemize}
\item Since $G^0$ is the neutral imaginary component of $\Phi_1$ and this doublet does not couple with fermions,
$G^0$ has no Yukawa interactions.
\item The quark masses depend only on $v_2$ therefore there are no fermionic contributions to the derivatives of the potential
with respect to $v_1$.
\item Another consequence is that the tensors $y^{IJG^0}$ are automatically zero, which implies that the
contributions to the self-energies, $\Pi^F_{G^0G^0}$ and  $\Pi^F_{AG^0}$, are also zero.
\item As a consequence the masslessness of the neutral Goldstone boson $G^0$ at one-loop is automatically confirmed.
\item The $y^{IJA}$ tensor becomes extremely simple, leading to no FCNC. The entries pertaining to the top quark, for instance, are given by
\beq
y^{56A}\,=\,y^{65A}\,=\, \frac{i\,m_t}{v_2}\,,
\eeq
and
\beq
M^{56} \,=\,M^{65}\,=\,m_t\,.
\eeq
\end{itemize}
For example, the one-loop contribution of the top quark to the $A$ mass will then be given by (using the definitions of
eqs.~\eqref{eq:Xizes} and~\eqref{eq:minF})
\bea
\frac{1}{16\pi^2}\,X^{top}_{AA} &=&
 \Pi^{top}_{AA} \,-\, \,\,\frac{(-12)}{2 v_2} \,\frac{\partial  m^2_t}{\partial
v_2}\,A(m^2_t) \nonumber \\
 & = &  \Pi^{top}_{AA}  \,+\,  12 \,\frac{m^2_t}{v_2^2} \,A(m^2_t) \,,
\eea
where the self-energy term is, according to~\eqref{eq:Piij},
\bea
\Pi^{top}_{AA} \,=\,2\times 3\times {\rm Re} \left[ y^{56A} y_{56A} \right] B_{FF} (m_t^2, m_t^2)
\,+\, 2\times 3 \times {\rm Re}  \left[ y^{56A} y^{56A} M_{56} M_{56}\right]
   B_{\bar{F}\bar{F}} (m_t^2, m_t^2) \,,
   \label{eq:PiAAt}
\eea
where the overall factor ``2" comes from the sum on the symmetric $y^{IJ}$ tensors indices, and ``3" accounts
for the colour of the quark. Then, from eqs.~\eqref{eq:BFF},~\eqref{eq:BFbFb} and~\eqref{eq:Bxxs0}, we obtain
\bea
B_{FF} (m_t^2, m_t^2) & = & -\,2\, m^2_t\,{\log}{\left(\frac{m^2_t}{\mu^2}\right)} \,-\,2\,A(m^2_t)\,, \nonumber \\
B_{\bar{F}\bar{F}} (m_t^2, m_t^2) & = &  -\,2 \,{\log}{\left(\frac{m^2_t}{\mu^2}\right)}\,.
\eea
Therefore, using the values shown above for $y^{56A}$ and $M_{56}$ in eq.~\eqref{eq:PiAAt} we find that
\beq
\Pi^{top}_{AA} \,=\,-\,12\,\frac{m^2_t}{v_2^2} \,A(m^2_t)\,,
\label{eq:pitop}
\eeq
which implies that $X^{top}_{AA} = 0$. The same conclusion will apply for the contributions for all other quarks.
Thus, in the Type-I $U(1)$ model the pseudoscalar remains massless at one-loop when that symmetry is spontaneously
broken, as expected by Goldstone's theorem. Since this demonstration can be made for each quark
separately, the conclusions we reached hold for models Type-II, X or Y, where no FCNC occurs in the model.

We tested all possible $U(1)$ Yukawa textures listed in ref.~\cite{Ferreira:2010ir} (to wit, those textures for the down and up
quarks shown in sections III C.1 to C.3 of that reference). In that paper, textures resulting from the application of abelian
global symmetries on the 2HDM lagrangian were obtained, with the demands that the symmetry-constrained Yukawa matrices
lead to six massive quarks, and a CKM matrix with no zero entries. These minimal requirements were not tested numerically (meaning,
the Yukawa textures were not shown to reproduce the correct values of the quark masses of entries of the CKM matrix). Several
textures were found to be possible only under application of a $Z_2$ symmetry -- thus not leading to a massless pseudoscalar --
or of a $Z_3$ symmetry, which we will treat in the next section. Unlike the flavour-preserving textures of Type-I (or II, X and Y),
which we showed analytically that lead to a massless pseudoscalar at the one-loop level, for the vast majority of the
$U(1)$ textures presented in~\cite{Ferreira:2010ir} the calculations of the self-energies can only be made numerically.
We use the procedure outlined above in section~\ref{sec:onelf}, generating random numbers for the Yukawa matrices' entries
and for the vevs~\footnote{We restricted ourselves to real entries for the Yukawa couplings, but there should be no differences
if complex entries were considered. }; diagonalizing numerically the resulting quark mass matrices, obtaining the right-handed and
left-handed rotation matrices defined in eq.~\eqref{eq:Mdu}; obtaining the contributions from the quarks to the one-loop
minimization conditions through the process explained in eqs.~\eqref{eq:derimp}, \eqref{eq:derFu} and~\eqref{eq:minF}; obtaining
the mass-basis Yukawa matrices through eq.~\eqref{eq:rotmat} and identifying the entries of the $y^{IJk}$ tensors following the
procedure outlined in eq.~\eqref{eq:ident}; and finally computing the quark contributions to the pseudoscalar mass matrix
following the formulae of eq.~\eqref{eq:Piij}.

The conclusion was the same for all the $U(1)$ Yukawa textures from~\cite{Ferreira:2010ir}: if the $U(1)$ symmetry
has been spontaneously broken, the one-loop pseudoscalar mass matrix yields two massless eigenvalues, even when including fermionic
contributions. This was expected, since it is exactly what Goldstone's theorem predicted. It may be seen as an explicit confirmation
of that theorem, but in the context of this paper it serves as a confirmation that our numerical methods of performing the mass
corrections at one-loop are correct.

\subsection{$Z_3$-symmetric Yukawa textures}

As we discussed in section~\ref{sec:mod}, it is possible to have Yukawa textures which are $Z_3$-symmetric -- those shown in
eq.~\eqref{eq:Yukz3} are a specific example, they can only be obtained if the angle $\theta$, defined in eq.~\eqref{eq:u1},
is equal to $2\pi/3$. As shown in eqs.~\eqref{eq:fases}, the several phases transforming left and right handed quark fields are also
constrained to be multiples of $\pm 2\pi/3$. An important point for what we will show below is the observation that the textures of the $\Gamma$
(or $\Delta$) matrices in eq.~\eqref{eq:Yukz3} require $\theta = 2\pi/3$ on their own, without the need to consider symmetry constraints
arising from the $\Delta$ ($\Gamma$) matrices. Other implementations of a $Z_3$ symmetry, such as the one studied in ref.~\cite{Ferreira:2011xc},
also require $\theta = 2\pi/3$, but to reach that conclusion one needs to analyze the restrictions arising from non-zero textures in
both the up and down Yukawa matrices, not just those of the up sector, or just those of the down one.

\begin{figure}[ht]
  \centering
\includegraphics[height=7cm,angle=0]{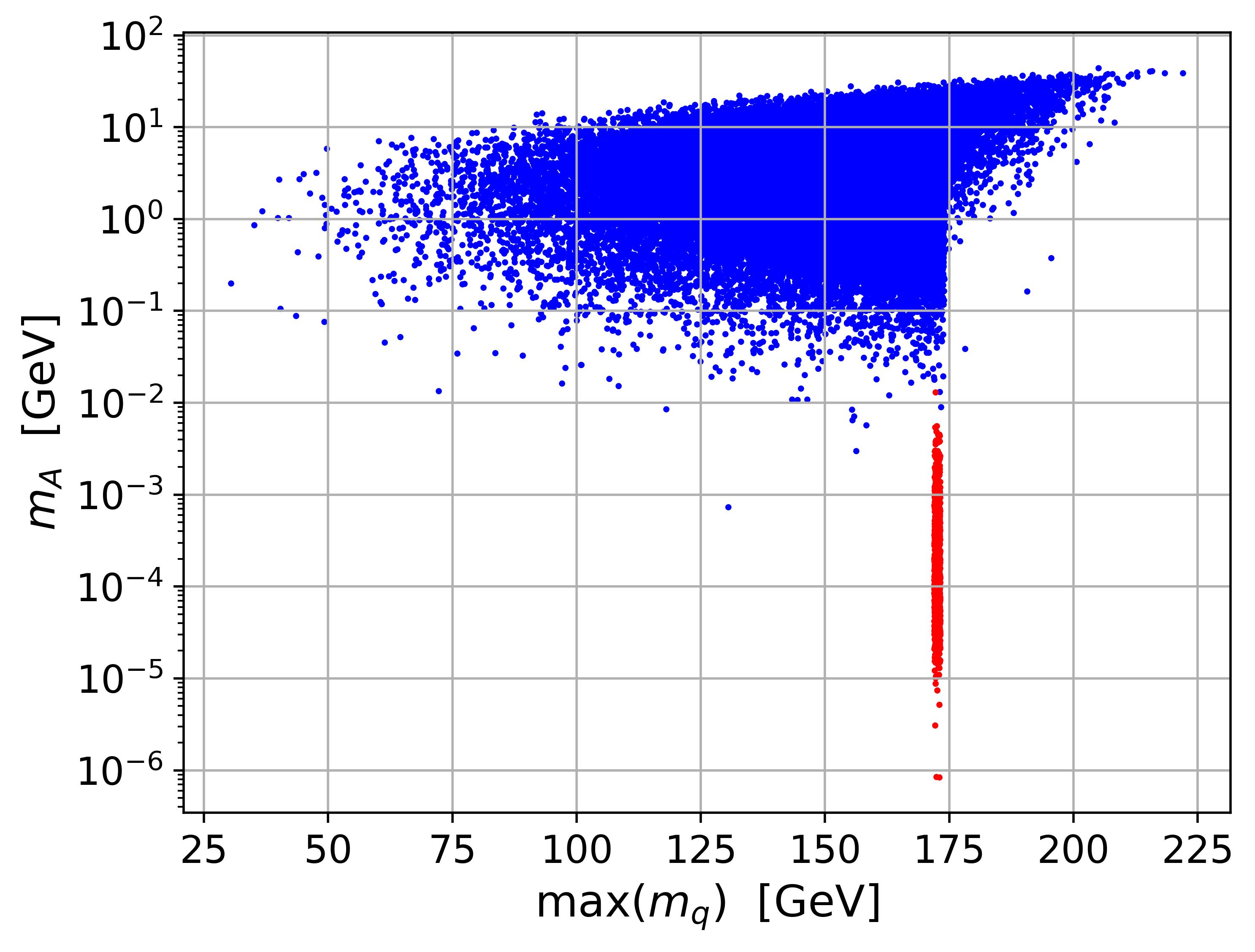}
  \caption{Pseudoscalar mass as a function of the highest quark mass obtained for the $Z_3$ Yukawa textures of eq.~\eqref{eq:Yukz3}. In red, points
  for which all six quark masses were fitted to their experimental values.}
\label{fig:mAmq}
\end{figure}

We performed a numerical analysis of the one-loop quark contributions to the pseudoscalar mass arising from the Yukawa textures of
eq.~\eqref{eq:Yukz3}. As we did for the $U(1)$ textures, we generated random numbers for the values of the Yukawa couplings (between
-1 and 1) and
for the vevs $v_1$ and $v_2$ such that $v_1^2+v_2^2 = v^2$ and computed the quark mass matrices, their eigenvalues and right-handed
and left-handed rotation matrices. We then calculated the quark contributions to the one-loop minimization conditions and to the
pseudoscalar self-energies. In all that follows we chose $\mu = 100$ GeV~\footnote{The physical results are independent of
this choice, though it is assumed that the value of $\mu$ is such that the logarithmic terms in the potential and self-energies
are small. For coherence, this also means that the values of all running quantities are being taken at the scale $\mu$.}.
The results are shown in figure~\ref{fig:mAmq},
where we show the pseudoscalar mass $m_A$ as a function of the highest quark mass obtained in the numerical procedure described above.
In the case of the blue points in the plot we did not require that the quark masses obtained are correct, but in a separate fit we generated red points
using a minimization procedure to find the
values of the Yukawas which, for a given set of vevs, correctly reproduce all six quark masses within 2-$\sigma$ intervals of their experimental values,
taken from~\cite{ParticleDataGroup:2024cfk}. We computed the self-energies at $s = 0$~\footnote{This is a good approximation~\cite{Ellis:1990nz,Ellis:1991zd},
and we confirmed its validity performing the iterative procedure mentioned earlier. We found that convergence to $s = m_A^2$ occurs after less than
7 iterations for most cases and that the final value of $m_A$ is usually at most 6\% different from the one computed with $s = 0$.}.
We verify that the neutral Goldstone boson
continues, as expected, to be massless, but the same cannot now be said for the pseudoscalar.

The most relevant conclusion from figure~\ref{fig:mAmq} is that the one-loop quark contributions force $m_A \neq 0$, thus confirming what is
expected from Goldstone's theorem: since
the lagrangian has a discrete $Z_3$ symmetry -- not a continuous $U(1)$ one -- then a spontaneous breaking of that symmetry
ought to not imply a massless pseudoscalar. That in the 2HDM with a $Z_3$ symmetry that occurs at tree-level is an accident,
arising from the fact that the scalar potential is identical for both $Z_3$ and $U(1)$ symmetries. But here we see how the
lagrangian with a discrete symmetry implies a massive pseudoscalar.

The second notable conclusion is that the pseudoscalar mass obtained is much smaller when all six quark masses obtained from the $Z_3$
Yukawa textures are in agreement with experimental values -- this, we believe, is a consequence of the hierarchy of quark masses, which
ranges from $m_u \simeq 2.2$ MeV to $m_t \simeq 173$ GeV. In order to obtain such diverse orders of magnitude in the quark masses with
vevs $v_1$ and $v_2$ ranging between 0 and 246 GeV, the entries of the $Z_3$ Yukawa matrices of eq.~\eqref{eq:Yukz3} must also have
significant differences in order of magnitude. What we have concluded from our parameter scan is that, in order to correctly reproduce
the known spectrum of up and down quark masses, at least one entry of the matrices $\Gamma$ and another in matrices $\Delta$ must be
several orders of magnitude (between two and five) below the remaining ones. It is easy to see that, if one (any one) of
the non-zero entries in the Yukawa matrices shown in eq.~\eqref{eq:Yukz3} is set to zero, the resulting theory is invariant under a
full $U(1)$ symmetry, no longer under a discrete $Z_3$ one. Thus, fitting all quark masses to their real values and therefore
reproducing the observed fermion hierarchy forces our $Z_3$ Yukawa matrices to numerically approach a $U(1)$ texture -- which, if exact,
would force $m_A = 0$. A corresponding reduction in the pseudoscalar mass is thus obtained.

\begin{figure}[t]
\begin{tabular}{cc}
\includegraphics[height=6cm,angle=0]{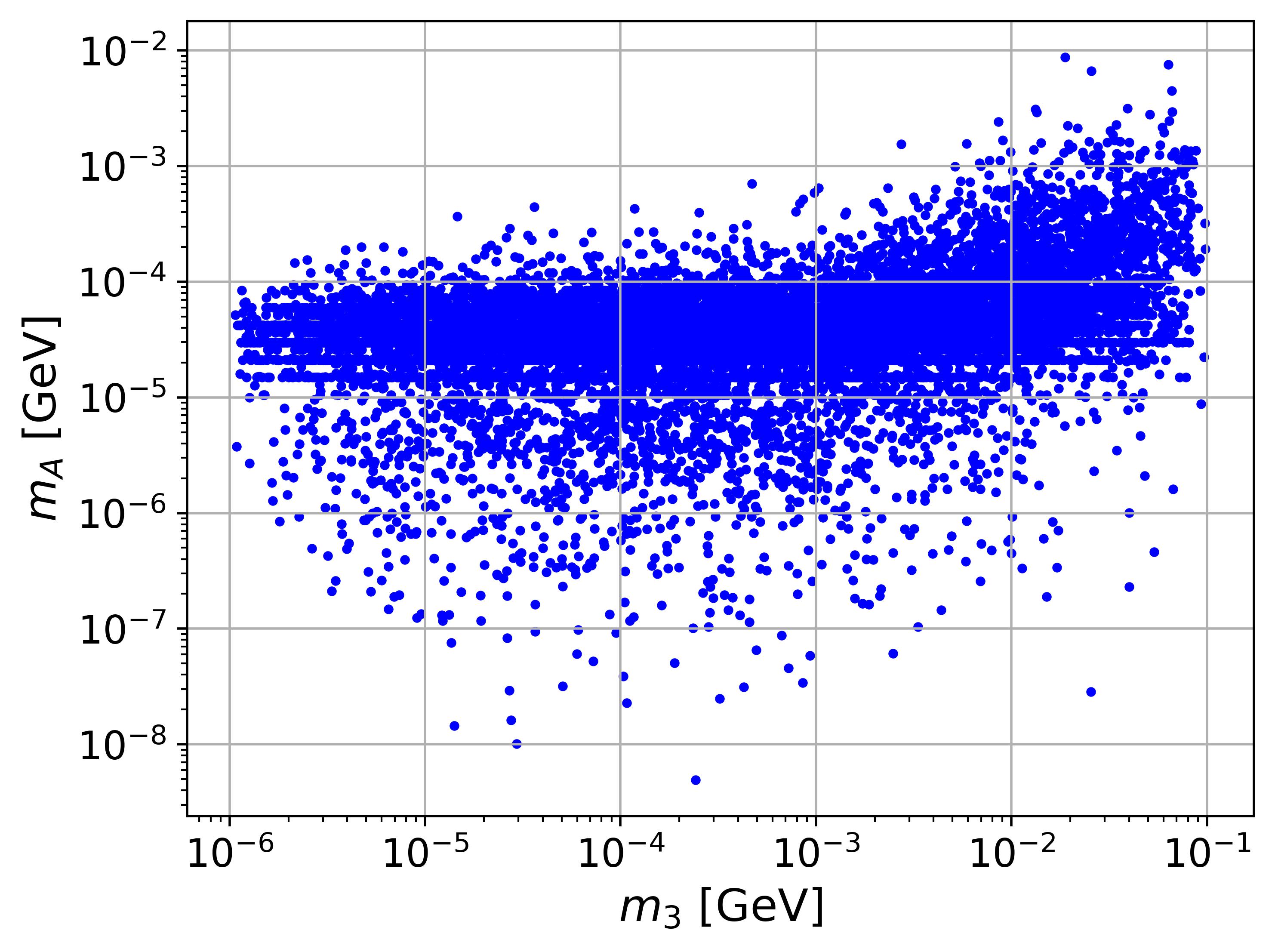}&
\includegraphics[height=6cm,angle=0]{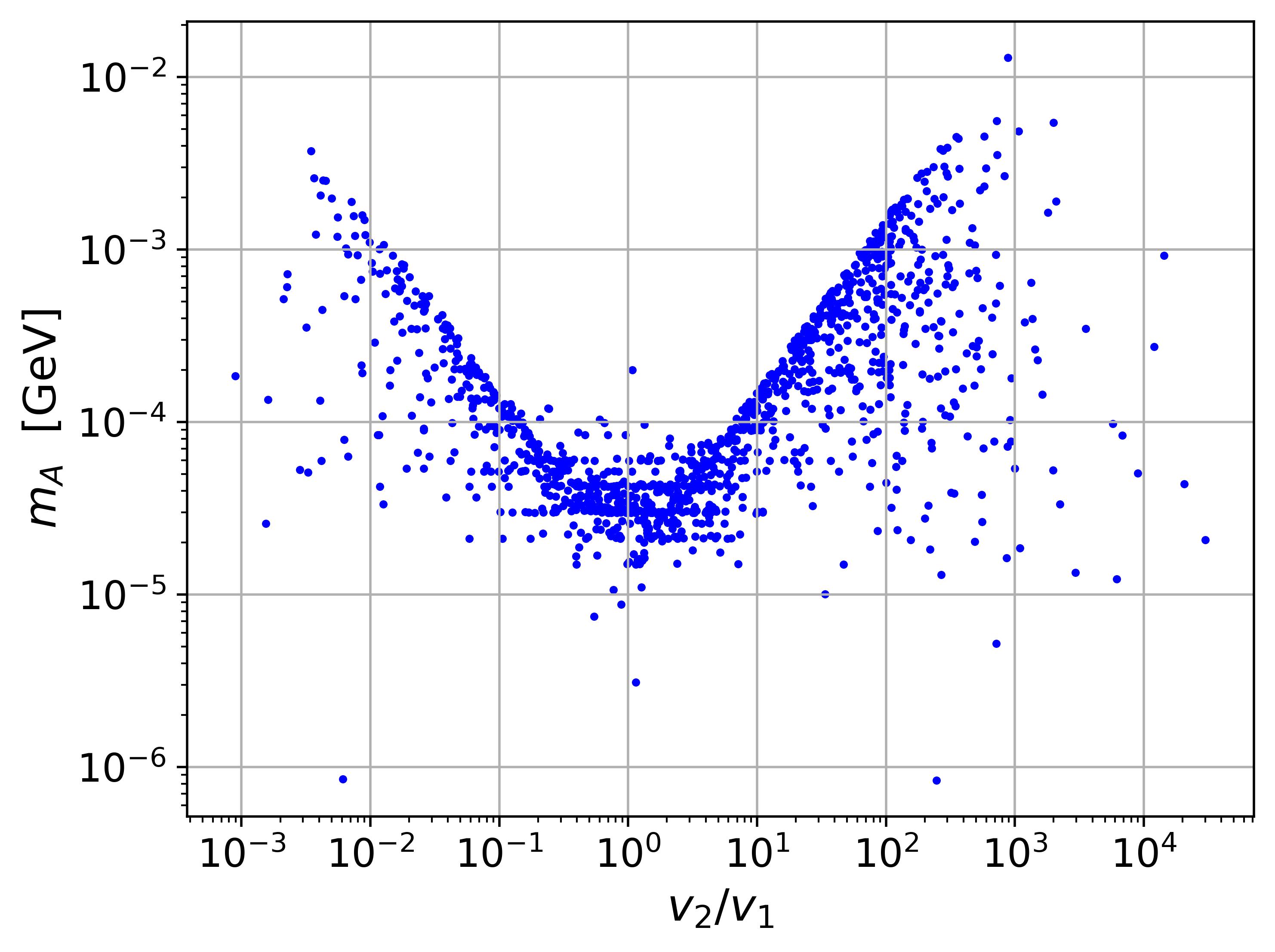}\\
 (a) & (b)
\end{tabular}
\caption{Scatter plots for the one-loop pseudoscalar mass as a function of
(a) Minimum value of the quark mass generated, fixing the highest mass to 10 GeV and allowing the remaining two to differ by
several orders of magnitude. Only up quarks considered.
(b) Vev ratio $v_2/v_1$, keeping $v_1^2+v_2^2 = v^2$. All six quark masses correctly fitted.
}
\label{fig:hie}
\end{figure}

To verify this, we performed two separate analysis, whose results are shown in figure~\ref{fig:hie}.
In plot~\ref{fig:hie} (a), we considered only the contributions from one type of quarks (in this case, the up ones) and fixed
one of their masses ($m_1$) to 10 GeV. The second mass ($m_2$) we allowed to vary from 0.01 to 1 GeV, and the
third mass $m_3$ was allowed to vary between $10^{-6}$ and $10^{-3}$ GeV. In this way we forced a very hierarchical quark mass spectrum.
The effect is clear:
as the fermion mass spectrum becomes more hierarchical (meaning, as the smallest quark mass becomes increasingly smaller),
the pseudoscalar mass is driven to smaller values. In figure~\ref{fig:hie} (b) we attempted a different fitting procedure,
randomly generating vevs $v_1$ and $v_2$ correctly reproducing electroweak symmetry breaking, thus satisfying
$v_1^2+v_2^2 = v^2 = (246 \mbox{ GeV})^2$ but including cases with vevs with significant differences in orders of
magnitude. In this way we were trying to ``shift" the hierarchy in quark masses to a hierarchy in vevs, which would allow for
all Yukawa couplings to be of the same order in magnitude, but the conclusion remains: it is not possible to reproduce
the experimental values of all six quark masses and not have at least one entry be several orders of magnitude smaller than the
rest, which makes the $Z_3$-invariant matrices be numerically similar
to $U(1)$-invariant ones, thus yielding a much smaller pseudoscalar mass.

This model provides a rather natural mechanism to generate a very low pseudoscalar mass -- $m_A$ would be small because it is actually
zero at leading order, its non-zero mass the result of radiative corrections. Indeed, we found a rough upper bound of $\sim 0.008$
GeV on the pseudoscalar mass in this model. Is such a low pseudoscalar mass excluded by experimental axion searches? We can use
the exclusion limits shown in figure 90.1 of~\cite{ParticleDataGroup:2024cfk} as a function of the pseudoscalar mass $m_A$ and its coupling to two
photons $g_{A\gamma\gamma}$. In the $Z_3$ model we are studying, the axion would have a diphoton decay through a fermion triangle loop. There
are very well known formulae for the width of such a decay (see for instance~\cite{Djouadi:2005gj,Barroso:2012wz}). Using the definitions
from~\cite{ParticleDataGroup:2024cfk} and~\cite{Barroso:2012wz}, we obtain
\beq
g_{A\gamma\gamma}\,=\, \frac{\alpha}{6\sqrt{2}\pi}\, \left| 4\frac{\left(\bar{\Delta}_2\right)_{33}}{m_t}\,A^A_{1/2}(\tau_t)
\,+\, \frac{\left(\bar{\Gamma}_2\right)_{33}}{m_b}\,A^A_{1/2}(\tau_b) \,+\, \cdots \right|\,,
\label{eq:gAphph}
\eeq
where $\tau_x = m_A^2/(4 m^2_x)$, $A^A_{1/2}(\tau)$ are the well-known pseudoscalar form factors and we explicitly show the top and bottom
contributions. With the range of pseudoscalar
masses which correctly fit the quark masses (see figure~\ref{fig:mAmq}) to very good approximation we have $\tau_t\simeq \tau_b\simeq 0$
and in that limit it is easy to obtain $A^A_{1/2}(0)\simeq 2$. For the parameter scan we performed, we obtain $g_{A\gamma\gamma}$
between roughly $10^{-9}$ and $10^{-3}$ GeV$^{-1}$ for pseudoscalar masses ranging from about $10^{2}$ to $8\times 10^6$ eV
and this parameter range is excluded by several astrophysical searches
(see~\cite{ParticleDataGroup:2024cfk} for details). A careful consideration of the assumptions undertaken in those
searches would be interesting, but falls outside the scope of the current work.

At this stage we should investigate the other $Z_3$-invariant Yukawa matrices presented in ref.~\cite{Ferreira:2010ir}. We did so,
and discovered that, out of all the $Z_3$ invariant textures shown in that paper (to wit, those contained in equations (89) to
(95)), the only one that did {\em not} yield a one-loop massless pseudoscalar is the one shown in our eq~\eqref{eq:Yukz3}
(eq. (92) of~\cite{Ferreira:2010ir}).  Though surprising at first, there is a good reason for this to occur: all the other
apparently $Z_3$ textures shown in that paper can actually be shown to be related by basis transformations to $U(1)$ invariant ones
(to be specific, to the textures shown in equations (57) or (60) of~\cite{Ferreira:2010ir}). Therefore they are not really $Z_3$ invariant,
but rather invariant under a continuous $U(1)$ symmetry, and as such the pseudoscalar remains massless at one-loop for such
Yukawa textures.

An even more interesting situation occurs with the $Z_3$ Yukawa textures used in ref.~\cite{Ferreira:2011xc},
which also lead to the pseudoscalar mass remaining equal to zero at the one-loop level. In that paper, the Yukawa matrices are
found to be
\bea
\Gamma_1 \, = \, \begin{pmatrix} 0 & 0 & 0 \\ 0 & 0 & \times \\ \times & \times & 0 \end{pmatrix}  & , &
\Gamma_2 \, = \, \begin{pmatrix} \times & \times & 0 \\ 0 & 0 & 0 \\ 0 & 0 & \times \end{pmatrix}\,, \nonumber \\
\Delta_1 \, = \, \begin{pmatrix} 0 & 0 & 0 \\ 0 & 0 & \times \\ \times & \times & 0 \end{pmatrix}  & , &
\Delta_2 \, = \, \begin{pmatrix} 0 & 0 & \times \\ \times & \times & 0 \\ 0 & 0 & 0 \end{pmatrix}  \,,
\label{eq:Yukz3l}
\eea
so that the lagrangian is invariant under the choice of phases (defined in eqs.~\eqref{eq:u1} and~\eqref{eq:phq})
\bea
\alpha_2 - \beta_3 \,=\, 0 &,& \alpha_1 - \beta_1 - \theta \,=\, 0 \,\nonumber \\
\alpha_3 - \beta_1 \,=\, 0 &,& \alpha_1 - \beta_2 - \theta \,=\, 0 \,\nonumber \\
\alpha_3 - \beta_2 \,=\, 0 &,& \alpha_3 - \beta_3 - \theta \,=\, 0 \,
\label{eq:abg}
\eea
for the $\Gamma$ matrices (and all zeros above should be interpreted as integer multiples of $2\pi$) and,
for the $\Delta$ matrices,
\bea
\alpha_2 - \gamma_3 \,=\, 0 &,& \alpha_1 - \gamma_3 + \theta \,=\, 0 \,\nonumber \\
\alpha_3 - \gamma_1 \,=\, 0 &,& \alpha_2 - \gamma_1 + \theta \,=\, 0 \,\nonumber \\
\alpha_3 - \gamma_2 \,=\, 0 &,& \alpha_2 - \gamma_2 + \theta \,=\, 0 \, .
\label{eq:agd}
\eea
The equations~\eqref{eq:abg} coming from the constraints on the $\Gamma$ matrices lead to two equations
on the $\alpha$ phases,
\beq
\alpha_1 - \alpha_3\,=\, \theta\;\;\; , \;\;\;
\alpha_3 - \alpha_2 \,=\, \theta \,+\, 2n\pi\,,
\label{eq:alg}
\eeq
for some integer $n$, which can be satisfied by any value of $\theta$ -- the $\Gamma$ matrix textures of eq.~\eqref{eq:Yukz3l} can therefore be
obtained by a generic $U(1)$ symmetry. Likewise, from the $\Delta$ matrix constraints in eq.~\eqref{eq:agd}, we obtain
\beq
\alpha_1 - \alpha_2\,=\, -\theta\;\;\; , \;\;\;
\alpha_2 - \alpha_3 \,=\, -\theta \,+\, 2m\pi\,,
\label{eq:ald}
\eeq
for any integer $m$, again equations which might be satisfied for any value of $\theta$ -- and thus the $\Delta$ matrix textures of eq.~\eqref{eq:Yukz3l} can
be obtained by a generic $U(1)$ symmetry. However, when we take eqs.~\eqref{eq:alg} and~\eqref{eq:ald} together, we immediately obtain that they
can only hold together if
\beq
3\theta \,=\, 2n\pi
\eeq
for some integer $n$, which can only be satisfied by $\theta = 2\pi/3 + 2n\pi$. This means that the only way to obtain the combined
$\Gamma$ and $\Delta$ textures of eq.~\eqref{eq:Yukz3l} is with a $Z_3$ symmetry.

Therefore the Yukawa textures considered in~\cite{Ferreira:2011xc} are
unquestionably the result of a $Z_3$ symmetry, but with a difference {\em vis a vis} those of eq.~\eqref{eq:u1} treated in this paper.
As we showed in eq.~\eqref{eq:fases}, the texture of the down-type Yukawa matrices $\Gamma$ alone forced the angle $\theta$ to be $2\pi/3$,
and the same can be shown for those $\Delta$ matrices. As shown above, however, the $\Gamma$ matrix textures used in ref.~\cite{Ferreira:2011xc},
could be reproduced by a $U(1)$ symmetry, and likewise for the corresponding $\Delta$ matrix textures. And this is relevant because, given
that we are considering neutral minima, the quark contributions to the self-energies in eq.~\eqref{eq:Piij} are computed separately for up and down quarks.
Thus, if the $\Gamma$/$\Delta$ matrices originating such contributions have textures which are analogous to those obtained from a $U(1)$ symmetry,
then they will give the same result as that of a continuous symmetry, preserving the masslessness of the pseudoscalar. Simply put, in the one-loop
formulae of eq.~\eqref{eq:Piij} the up-quark contributions ``don't know" about the down-quark ones and therefore, even though the
lagrangian of ref.~\cite{Ferreira:2011xc} is certainly invariant under a $Z_3$ symmetry through the interplay between up and down Yukawa matrices,
at one-loop even the fermionic contributions to the  self-energy will not render the pseudoscalar massive. It is to be expected that with higher order
corrections taken into account a non-zero mass contribution from the fermion sector will appear in this model, but that cannot occur at one-loop.
It may be an analogous situation to the ``propagation" of CP violation from the Yukawa sector to the rest of the theory in the
SM~\cite{Ellis:1978hq}.

\section{Conclusions}
\label{sec:conc}

Spontaneous breaking of a continuous symmetry yields massless scalars. This is the well-known Goldstone theorem, a crucial part
of understanding how the electroweak gauge bosons acquire their masses and longitudinal polarizations in the Higgs mechanism.
In the 2HDM, global symmetries are frequently considered to increase the model's predictive power, or to produce interesting
phenomenology. One such symmetry is the continuous Peccei-Quinn $U(1)$. When both doublets acquire a vev electroweak symmetry breaking
occurs and three Goldstone bosons appear, as usual -- but an additional massless pseudoscalar is produced, as expected.
Likewise, when imposing a $Z_3$ symmetry on the 2HDM, a massless pseudoscalar also appears at tree-level, despite that symmetry being a
discrete one. This is due to the fact that, the 2HDM potential having at most terms quartic in the scalar fields, imposing
a discrete $Z_3$ symmetry leads to a scalar sector with an accidental continuous $U(1)$. In fact, any $Z_N$ symmetry with
$N\geq 3$ leads to the same $U(1)$-invariant potential. While imposing these different symmetries leads to the same
scalar and gauge sectors, they can lead to different fermion sectors. Indeed, there are many different ways of extending
both the $U(1)$ and $Z_3$ symmetries to the fermion sector, with considerably different phenomenologies. Yukawa interactions
may distinguish between discrete and continuous symmetries, and one should expect that in a lagrangian invariant under
a discrete $Z_3$ symmetry the pseudoscalar mass will acquire loop corrections stemming from fermions -- but remain massless for any
$U(1)$-invariant model.

We tested this assumption by computing the one-loop self-energies to the pseudoscalar mass matrix, showing analytically that, for both
$U(1)$ and $Z_3$ invariant 2HDMs, they yielded two massless particles when only the contributions from the scalar and gauge sectors
were taken into account. This was to be expected, since those two sectors of the model cannot distinguish between both symmetries.
We then computed the one-loop quark contributions to the pseudoscalar mass matrix for all $U(1)$-invariant Yukawa
textures in the exhaustive list provided in ref.~\cite{Ferreira:2010ir}, and showed that for all of them the pseudoscalar
remained massless at one-loop, as foreseen by Goldstone's theorem. Repeating that calculation for a specific $Z_3$ Yukawa
texture (eq. (95) of ref.~\cite{Ferreira:2010ir}, eq.~\eqref{eq:Yukz3} of this paper) we then found a non-zero pseudoscalar mass
(the neutral Goldstone from electroweak symmetry breaking remaining massless, of course). Thus we confirmed that imposing a discrete
symmetry on the 2HDM lagrangian did not lead to a massless scalar -- the accidental tree-level symmetry of the scalar potential was
not verified when one-loop fermion corrections were taken into account. We also found that several symmetries classified as $Z_3$
in~\cite{Ferreira:2010ir} were indeed basis changes from $U(1)$ symmetries, thus producing massless axions even at one-loop.

Two unexpected things were revealed in our calculation, however. The first is that, though Yukawa interactions give rise to a one-loop
pseudoscalar mass $m_A$, the value of $m_A$ is highly, and non-intuitively, dependent on the quark mass spectrum. Indeed, allowing for
generic quark masses all of the same order yields one-loop fermion masses as high as $\sim$ 30 GeV, as we see in figure~\ref{fig:mAmq}
for a maximum quark mass of $\sim$ 200 GeV. We can understand these values by performing a quick estimate, approximating the
fermion contributions to the scalar masses by the second derivative of the potential with respect to the vevs; we take $A(x)\simeq x$
in eq.~\eqref{eq:minF} and the maximum value $v_2$ can take, so that
\beq
m_A \,\simeq \, \sqrt{\frac{\partial^2 V_1^F}{\partial v_2^2} } \,=\,\frac{1}{8\pi} \sqrt{12 \frac{m_q^4}{v^2} }\, \simeq 22.4 \,\,\rm{ GeV}\,.
\eeq
However, when adjusting all six quark masses to their know values, we see, in the red points of figure~\ref{fig:mAmq} that the
pseudoscalar mass becomes much smaller, between $10^{-6}$ and $10^{-2}$ GeV in our fit. We interpreted this as a consequence of
using the Yukawa matrices of eq.~\eqref{eq:Yukz3} to
reproduce the strong hierarchy of the quark mass spectrum: to obtain such different masses, some of the entries of the Yukawa matrices must
be much smaller than others. But when any of the entries of the matrices in eq.~\eqref{eq:Yukz3} are zero, the lagrangian ends up invariant
under a continuous $U(1)$ symmetry, which leads to $m_A = 0$. The fermion mass hierarchy thus leads to an {\em approximate} continuous
symmetry, thus reducing substantially the order of magnitude of the pseudoscalar mass. The model therefore provides a natural way
of generating very small ($\sim$ MeV and lower) axion masses. Since these axions interact with charged fermions they have a loop-induced
diphoton decay. The coupling strength of the pseudoscalar-photon interactions determined from those interactions makes the parameter
space studied here already excluded by experimental axion searches, though a deeper analysis is warranted (for instance, some axion searches
assume the particle behaves like Dark Matter, which may not be a reasonable assumption for the model under discussion).

The second unexpected result was finding a massless pseudoscalar for certain $Z_3$ Yukawa textures (those of ref.~\cite{Ferreira:2011xc},
our eq.~\eqref{eq:Yukz3l}).
This is a consequence of the fact that the textures in the down/up Yukawa matrices can be reproduced separately with a $U(1)$
symmetry -- it is only when one tries to force {\em both} down and up Yukawa matrices to have the forms shown in eq.~\eqref{eq:Yukz3l}
that one finds that that is only possible for a $Z_3$ symmetry, not for a generic $U(1)$. But the one-loop pseudoscalar fermionic
self-energies have separate contributions from the up and down quarks -- hence, if the Yukawa matrices contributing to those self
energies have identical textures to others resulting from a $U(1)$ symmetry, the net contribution to pseudoscalar masses
will be zero. One expects that Goldstone's theorem still is valid, but only higher order corrections, presumably
involving diagrams with both up  and down quarks, will eventually result in $m_A \neq 0$. We might then expect that the model
studied in~\cite{Ferreira:2011xc}, if devoid of any soft breaking terms, would produce in a natural way much smaller values for
the pseudoscalar mass. Thus, like the consequences of CP violation in the SM on the rest of the lagrangian, the impact of a discrete
symmetry in the Yukawa sector may  only reflect itself in the pseudoscalar mass at very high orders. On a side note, the work
of refs.~\cite{Braathen:2016cqe,Braathen:2017izn} should be especially appropriate for these questions, as their authors have
performed two-loop scalar mass calculations for generic renormalizable theories, using 
SARAH implementations~\cite{Staub:2008uz,Staub:2009bi,Staub:2010jh,Staub:2012pb,Staub:2013tta}. The tools therein developped should
allow for a verification of the results of this work, and a possible way of confirming the two-loop conjecture for the 
textures of~\eqref{eq:Yukz3l}.

The conclusions reached in this work are valid for any theory with discrete symmetries which lead to accidental continuous ones
in the scalar potential at tree-level. Goldstone's theorem assures us that, if the model's lagrangian distinguishes between a
continuous and a discrete symmetry
in the Yukawa sector, there will be fermionic contributions that will make the pseudoscalar massive. However, it is not guaranteed
that that will occur already at the one-loop level. The two $Z_3$ symmetries compared in this paper showed that, if up/down
Yukawa matrices are formally identical to those obtained from a $U(1)$ symmetry, $m_A$ will remain zero at east at the one-loop level.
An interesting question becomes therefore: other than the textures of eq.~\eqref{eq:Yukz3}, can there be any other $Z_3$-symmetric
Yukawas that alread yield a non-zero pseudoscalar mass at one-loop? The analysis of ref.~\cite{Ferreira:2010ir} was very thorough, but
it did not seem to include the textures of ref.~\cite{Ferreira:2011xc}, for instance. Finally, what applications of interest might there
be in these calculations? Though our motivation was essentially a theoretical one, the fact that we found low axion masses arising from
symmetries may well be of interest in exploring the allowed parameter space for those particles shown in ref.~\cite{ParticleDataGroup:2024cfk}.
In particular, it would be interesting to explore models with discrete symmetries for which the pseudoscalar mass is loop-generated and
thus naturally small but where the axion couplings to diphotons is also suppressed (unlike the model we studied here). This would require
a suppression of the axion-fermion couplings contributing to $g_{A\gamma\gamma}$ (see eq.~\eqref{eq:gAphph}), which might be possible in
a version of the so-called CP3 model.

\section*{Acknowledgments}

We would like to thank Johannes Braathen and Apostolos Pilaftsis for their constructive and helpful remarks. 
This work is supported
by \textit{Funda\c c\~ao para a Ci\^encia e a Tecnologia} (FCT)
through contracts  \linebreak 
UIDB/00618/2020, UIDP/00618/2020, CERN/FIS-PAR/0025/2021 and 2024.03328.CERN.

\appendix
\section{Coupling and integral definitions}
\label{ap:def}

Below we define the several integral functions used in the self-energy calculations. Our definitions follow the
conventions of~\cite{Martin:2003qz,Martin:2003it}:
\bea
A(x) &=& x\,\left[{\log}{\left(\frac{x}{\mu^2}\right)}\,-\,1\right]\,,
\label{eq:A} \\
A_G(x) &=& 4 A(x) -  {\mathcal L}_x [x A(x)]\,,
\label{eq:AG} \\
 B(x,y) &=&
2 - r_{sxy} \, {\log}{\left(\frac{x}{\mu^2}\right)} - t_{syx} \, {\log}{\left(\frac{y}{\mu^2}\right)} +
\frac{\Delta^{1/2}_{sxy}}{s}\, \log (t_{xys}) \,,
\label{eq:B} \\
B_{SG} (x,y) & = & (2x-y+2s) B(x,y) + A(x) - 2 A(y) \nonumber \\
 & & + {\mathcal L}_y [(x+y-s) A(y) - (x-s)^2 B(x,y)]\,,
\label{eq:BSV}  \\
B_{GG} (x, y) & = &  -\frac{7}{2} B(x,y)
+ \frac{1}{2} {\mathcal L}_x [x B(x,y)]
+ \frac{1}{2} {\mathcal L}_y [y B(x,y)]\,,
\nonumber \\ & &
+ \frac{1}{4} {\mathcal L}_x {\mathcal L}_y \bigl \lbrace
x A(y) + y A(x)
+[ 2 s (x+y) -x^2 -y^2 -s^2 ] B(x,y)
\bigr \rbrace
\label{} \\
B_{FF} (x,y) & = &  (x+y-s) B(x,y) - A(x)-A(y)\,,
\label{eq:BFF} \\
B_{\bar{F}\bar{F}} (x, y) & = & 2 B(x,y)\,,
\label{eq:BFbFb}
\eea
where $A_G$ and $B_{GG}$ are presented for the $\overline{DR}$ renormalization scheme chosen and the following definitions were used:
\bea
\Delta_{xyz} &=& x^2 + y^2 + z^2 - 2 x y -2 x z -2 y z \,, \nonumber \\
t_{abc} &=& \frac{a+b-c + \Delta_{abc}^{1/2}}{2a}\,, \nonumber \\
r_{abc} &=& \frac{a+b-c - \Delta_{abc}^{1/2}}{2a}\,.
\eea
Also, we have
\begin{eqnarray}
{\mathcal L}_x f(x)\, \equiv\, \frac{f(x) - f( \xi x)}{x}\,,
\end{eqnarray}
where $\xi=0\,,1$ corresponds to the Landau/Feynman gauges.
At $s = 0$ we have the following simplified expressions:
\bea
B(x,y) &=& \frac{A(x) - A(y)}{y - x} \,,
\label{eq:Bs0} \\
B(x,x) &=& -\,{\log}{\left(\frac{x}{\mu^2}\right)} \,,
\label{eq:Bxxs0} \\
B_{SG}(x,y) &=& 0 \,,
\label{eq:BSV0}
\eea
where the $B_{SG}$ condition was obtained in the Landau gauge.

Here we present the relevant non-zero couplings used in our calculations (the ones not shown may be obtained from permutations of indices). The scalar trilinear couplings defined in eq.~\eqref{eq:LS} are:
\bea
\lambda^{G^0 H^+ G^-} &=& -\lambda^{G^0 G^+ H^-}\,=\,-\,\frac{1}{2} v_2 \lambda_4\,, \nonumber\\
\lambda^{A H^+ G^-} &=& -\lambda^{A G^+ H^-}\,=\,\frac{1}{2} v_1 \lambda_4\,,\nonumber\\
\lambda^{G^0 G^0 h} &=& v_1 \lambda_1 c_\alpha - v_2 \lambda_{34} s_\alpha\,,\nonumber\\
\lambda^{G^0 G^0 H} &=&  v_2 \lambda_{34} c_\alpha + v_1 \lambda_1 s_\alpha\,,\nonumber\\
\lambda^{A A h} &=& v_1 \lambda_{34} c_\alpha - v_2 \lambda_{2} s_\alpha\,, \nonumber\\
\lambda^{A A H} &=&  v_2 \lambda_2 c_\alpha + v_1 \lambda_{34} s_\alpha\,,
\eea
and the quartic scalar couplings are:
\bea
\lambda^{G^0 A H^+ H^+}  &=&  \lambda^{G^0 A H^- H^-}\,=\,-\,\lambda^{G^0 A G^+ G^+} \,=\,-\,\lambda^{G^0 A G^- G^-}\,=\, -\lambda_4 c_\beta s_\beta\,,\nonumber\\
\lambda^{G^0 A H^+ G^+}  &=&  \lambda^{G^0 A H^- G^-}\,=\, \frac{1}{2} \lambda_4 ( c_\beta^2 - s_\beta^2)\,,\nonumber\\
\lambda^{G^0 G^0 H^+ H^+}  &=&  \lambda^{G^0 G^0 H^- H^-}\,=\, \lambda_1 c_\beta^2 + \lambda_3 s_\beta^2\,,\nonumber\\
\lambda^{G^0 G^0 G^+ G^+}  &=& \lambda^{G^0 G^0 G^- G^-}\,=\, \lambda_3 c_\beta^2 + \lambda_1 s_\beta^2\,,\nonumber\\
\lambda^{G^0 G^0 H^+ G^+}  &=&  \lambda^{G^0 G^0 H^- G^-}\,=\, (\lambda_1-\lambda_3) c_\beta s_\beta\,,\nonumber\\
\lambda^{AA H^+ H^+}  &=& \lambda^{AA H^- H^-}\,=\, \lambda_3 c_\beta^2 + \lambda_2 s_\beta^2\,,\nonumber\\
\lambda^{AA G^+ G^+}  &=&  \lambda^{AA G^- G^-}\,=\, \lambda_2 c_\beta^2 + \lambda_3 s_\beta^2\,,\nonumber\\
\lambda^{AA H^+ G^+}  &=&  \lambda^{AA H^- G^-}\,=\, (\lambda_3-\lambda_2) c_\beta s_\beta\,,\nonumber\\
\lambda^{G^0 G^0 h h} &=&   \lambda_1 c_\alpha^2 + \lambda_{34} s_\alpha^2\,,\nonumber\\
 \lambda^{G^0 G^0 H H} &=&  \lambda_{34} c_\alpha^2 + \lambda_{1} s_\alpha^2\,,\nonumber\\
\lambda^{G^0 G^0 h H}  &=&  (\lambda_1-\lambda_{34}) c_\alpha s_\alpha\,,\nonumber\\
\lambda^{AAhh} &=& \lambda_{34} c_\alpha^2 + \lambda_2 s_\alpha^2\,,\nonumber\\
 \lambda^{AAHH} &=& \lambda_{2} c_\alpha^2 + \lambda_{34} s_\alpha^2\,,\nonumber\\
\lambda^{AAhH}  &=& (\lambda_{34}-\lambda_2) c_\alpha s_\alpha\,,\nonumber\\
 \lambda^{G^0 G^0 G^0 G^0}  &=&  3\lambda_1\,,\nonumber\\
 \lambda^{AAAA} &=& 3\lambda_2\,, \nonumber\\
 \lambda^{G^0 G^0 AA} &=& \lambda_{34}\,.
\eea

Likewise, the non-zero trilinear gauge couplings in eq.~\eqref{eq:LSG} are:
\bea
 g^{W^+ G^0 H^+} &=& g^{W^- G^0 H^+}\,=\,i\,g^{W^+ G^0 H^-}\,=\,-i\,g^{W^- G^0 H^-}\,=\,\frac{m_W}{v\sqrt{2}}\,c_\beta\,, \nonumber\\
 g^{W^+ G^0 G^+} &=& g^{W^- G^0 G^+}\,=\,i\,g^{W^+ G^0 G^-}\,=\,-i\,g^{W^- G^0 G^-}\,=\,\frac{m_W}{v\sqrt{2}}\,s_\beta\,,\nonumber\\
 g^{W^+ A H^+} &=& g^{W^- A H^+}\,=\,i\,g^{W^+ A H^-}\,=\,-i\,g^{W^- A H^-}\,=\, -\,\frac{m_W}{v\sqrt{2}}\,s_\beta\,,\nonumber\\
 g^{W^+ A G^+} &=& g^{W^- A G^+}\,=\,i\,g^{W^+ A G^-}\,=\,-i\,g^{W^- A G^-}\,=\,\frac{m_W}{v\sqrt{2}}\,c_\beta\,,\nonumber\\
 g^{Z G^0 h} &=& g^{Z A H}\,=\, -\, \frac{m_Z}{v}\,c_\alpha\,,\nonumber\\
 g^{Z G^0 H} &=& -g^{Z A h}\,=\,-\, \frac{m_Z}{v}\,s_\alpha\,.
\eea
It should be noted that contrarily to the scalar couplings above, $g^{aij}$ are not symmetric under an interchange of scalar indices (see eq.~\eqref{eq:LSG}), and
$g^{ab G^0} = g^{abA}=0$ for all gauge fields $a$, $b$. The quartic couplings are:
\bea
& g^{W^+ W^- G^0 G^0}\,=\,g^{W^+ W^- AA}\,=\,\frac{2\, m_W^2}{v^2}\,,\nonumber\\
& g^{Z Z G^0 G^0}\,=\,g^{Z Z A A}\,=\,\frac{2\, m_Z^2}{v^2}\,.
\eea

Finally, for the fermion couplings of eq.~\eqref{eq:LF} we have:
\bea
y^{1\,2\,G^0} = \frac{i}{\sqrt{2}}\,\left(\bar{\Delta}_1^\dagger\right)_{11}\,, \quad\quad
& y^{1\,4\,G^0} = \frac{i}{\sqrt{2}}\,\left(\bar{\Delta}_1^\dagger\right)_{12}\,, & \quad\quad
y^{1\,6\,G^0} = \frac{i}{\sqrt{2}}\,\left(\bar{\Delta}_1^\dagger\right)_{13}\,, \nonumber\\
y^{2\,3\,G^0} = \frac{i}{\sqrt{2}}\,\left(\bar{\Delta}_1^\dagger\right)_{21}\,, \quad\quad
& y^{2\,5\,G^0} = \frac{i}{\sqrt{2}}\,\left(\bar{\Delta}_1^\dagger\right)_{31}\,, & \quad\quad
y^{3\,4\,G^0} = \frac{i}{\sqrt{2}}\,\left(\bar{\Delta}_1^\dagger\right)_{22}\,, \nonumber\\
y^{3\,6\,G^0} = \frac{i}{\sqrt{2}}\,\left(\bar{\Delta}_1^\dagger\right)_{23}\,, \quad\quad
& y^{4\,5\,G^0} = \frac{i}{\sqrt{2}}\,\left(\bar{\Delta}_1^\dagger\right)_{32}\,, & \quad\quad
y^{5\,6\,G^0} = \frac{i}{\sqrt{2}}\,\left(\bar{\Delta}_1^\dagger\right)_{33}\,, \nonumber\\
y^{7\,8\,G^0} = \frac{i}{\sqrt{2}}\,\left(\bar{\Gamma}_1^\dagger\right)_{11}\,, \quad\quad
& y^{7\,10\,G^0} = \frac{i}{\sqrt{2}}\,\left(\bar{\Gamma}_1^\dagger\right)_{12}\,, & \quad\quad
y^{7\,12\,G^0} = \frac{i}{\sqrt{2}}\,\left(\bar{\Gamma}_1^\dagger\right)_{13}\,, \nonumber\\
y^{8\,9\,G^0} = \frac{i}{\sqrt{2}}\,\left(\bar{\Gamma}_1^\dagger\right)_{21}\,, \quad\quad
& y^{6\,11\,G^0} = \frac{i}{\sqrt{2}}\,\left(\bar{\Gamma}_1^\dagger\right)_{31}\,, & \quad\quad
y^{9\,10\,G^0} = \frac{i}{\sqrt{2}}\,\left(\bar{\Gamma}_1^\dagger\right)_{22}\,, \nonumber\\
y^{9\,12\,G^0} = \frac{i}{\sqrt{2}}\,\left(\bar{\Gamma}_1^\dagger\right)_{23}\,, \quad\quad
& y^{10\,11\,G^0} = \frac{i}{\sqrt{2}}\,\left(\bar{\Gamma}_1^\dagger\right)_{32}\,, & \quad\quad
y^{11\,12\,G^0} = \frac{i}{\sqrt{2}}\,\left(\bar{\Gamma}_1^\dagger\right)_{33}\,. 
\eea
The couplings $y^{IJA}$ can be obtained from these by replacing $\Delta_1$ ($\Gamma_1$) by $\Delta_2$ ($\Gamma_2$). The coefficients $M^{IJ}$, defined in eq.~\eqref{eq:MIJ}, are the masses of the six quarks, so we have:
\bea
M^{1\,2} = m_u\,, \quad\quad & M^{3\,4} = m_c\,, & \quad\quad M^{5\,6} = m_t\,, \nonumber\\
M^{7\,8} = m_d\,, \quad\quad & M^{9\,10} = m_s\,, & \quad\quad M^{11\,12} = m_b\,.
\eea

{\footnotesize
\bibliographystyle{utphys}
\bibliography{biblio}
}

\end{document}